\documentclass{IEEEtran}
\usepackage{cite}
\usepackage{amsthm}
\usepackage{amsmath,amssymb,amsfonts}
\usepackage{geometry}
\usepackage{graphicx}
\usepackage{textcomp}
\usepackage{xcolor}
\usepackage{longtable}
\usepackage{bm}
\usepackage{graphicx,epstopdf,psfrag,url,cite,xcolor,multirow,float}
\usepackage{subfloat}
\usepackage{amsmath}
\usepackage{bbm}
\usepackage{algorithm}
\usepackage{algorithmic}
\usepackage{array}
\usepackage{tabularx}
\usepackage{booktabs}

\usepackage{times}
\usepackage{subfigure}
\usepackage{arydshln}
\begin{document}

% \title{Fast and Unified Learning for RIS-assisted Multi-User Semantic Communications}
\title{Learning to Optimize Joint Source and RIS-assisted Channel Encoding for Multi-User Semantic Communication Systems}

\author{Haidong Wang, Songhan Zhao, Bo Gu, Shimin Gong, Hongyang Du, and Ping Wang
\thanks{Haidong Wang, Songhan Zhao, Bo Gu, and Shimin Gong are with the School of Intelligent Systems Engineering, Sun Yat-sen University, China (e-mail: wanghd7@mail2.sysu.edu.cn; zhaosh55@mail2.sysu.edu.cn; gubo@mail.sysu.edu.cn; gongshm5@mail.sysu.edu.cn).
%Nguyen Cong Luong is with Faculty of Computer Science, Phenikaa University, Vietnam (email: luong.nguyencong@phenikaa-uni.edu.vn).
Hongyang Du is with the Department of Electrical and Computer Engineering, University of Hong Kong, HongKong, SAR, China (email: duhy@hku.hk). Ping Wang is with
the Department of Electrical Engineering and Computer Science, Lassonde School of Engineering, York University, Canada (e-mail: pingw@yorku.ca). % Changyan Yi is with the College of Computer Science and Technology, Nanjing University of Aeronautics and Astronautics, China (e-mail: changyan.yi@nuaa.edu.cn).
}

}

\maketitle
\begin{abstract}

In this paper, we explore a joint source and reconfigurable intelligent surface (RIS)-assisted channel encoding (JSRE) framework for multi-user semantic communications, where a deep neural network (DNN) extracts semantic features for all users and the RIS provides channel orthogonality, enabling a unified semantic encoding-decoding design. We aim to maximize the overall energy efficiency of semantic communications across all users by jointly optimizing the user scheduling, the RIS's phase shifts, and the semantic compression ratio. Although this joint optimization problem can be addressed using conventional deep reinforcement learning (DRL) methods, evaluating semantic similarity typically relies on extensive real environment interactions, which can incur heavy computational overhead during training. To address this challenge, we propose a truncated DRL (T-DRL) framework, where a DNN-based semantic similarity estimator is developed to rapidly estimate the similarity score. Moreover, the user scheduling strategy is tightly coupled with the semantic model configuration. To exploit this relationship, we further propose a semantic model caching mechanism that stores and reuses fine-tuned semantic models corresponding to different scheduling decisions. A Transformer-based actor network is employed within the DRL framework to dynamically generate action space conditioned on the current caching state. This avoids redundant retraining and further accelerates the convergence of the learning process. Numerical results demonstrate that the proposed JSRE framework significantly improves the system energy efficiency compared with the baseline methods. By training fewer semantic models, the proposed T-DRL framework significantly enhances the learning efficiency.

\end{abstract}

\begin{IEEEkeywords}
Semantic communications, RIS, deep reinforcement learning, source-channel encoding.
\end{IEEEkeywords}
\IEEEpeerreviewmaketitle
\let\thefootnote\relax\footnotetext{
}

\section{Introduction}
Driven by emerging applications such as the Internet of Things (IoT), autonomous driving, and extended reality, the exponential growth of wireless data traffic is placing unprecedented strain on conventional communication systems~\cite{11152698}. Semantic communication has recently emerged as a promising paradigm shift, prioritizing the transmission of essential meaning or intent over raw data symbols~\cite{Xiao-twc2024}. By focusing on semantic relevance, this approach offers significant potential to eliminate redundancy and improve spectral efficiency, particularly in resource-constrained wireless networks~\cite{Cai2025Importance}. Substantial research efforts have explored the integration of semantic communication with multi-access schemes, including non-orthogonal multiple access (NOMA)~\cite{Li2023NOMA-Semantic} and rate-splitting multiple access (RSMA)~\cite{Cheng2023RSMA-semantic-interest_}, both of which have demonstrated notable performance gains. However, these methods typically rely on sophisticated channel coding techniques for multi-user decoding, leading to increasingly complex receiver architectures and computational overhead that scales with the number of users~\cite{Zhang2023DeepMA,Lin2025AIAirMA}. Alternative approaches such as space division multiple access (SDMA) experiences degraded spatial separability with dense user deployments~\cite{Sherif2024IRS-NOMA,Singh2025HARIS}, while code division multiple access (CDMA) suffers from reduced spectral efficiency owing to the need for intricate coding schemes to suppress multi-user interference~\cite{Singh2025HARIS}. These limitations in achieving channel orthogonality highlight a promising opportunity to achieve a more efficient multi-user access scheme from a complementary perspective by jointly exploiting users' orthogonality in both channel and semantic feature domains.
\subsection{Motivations and Challenges}
Reconfigurable intelligent surfaces (RISs) have attracted significant attention for their ability to actively reshape the wireless propagation environment. An RIS consists of a large array of passive reflecting elements, each capable of applying a programmable phase shift to incident signals~\cite{Li2025IRS4jamming}. By dynamically coordinating these phase shifts, RISs have been shown to enhance desired signal reception, suppress interference, and thereby improve overall system performance~\cite{Ding2025DL-IRS,Sobhi-Givi2025DRL-IRS}. Consequently, RISs offer a promising way to strengthen channel orthogonality, facilitating efficient multi-user simultaneous transmissions. When the spatial orthogonality dynamically configured by the RIS is strategically combined with semantic orthogonality in the feature space, a more flexible and robust multi-access paradigm emerges through their complementary control. Specifically, when users are spatially separated and exhibit favorable channel orthogonality, they can achieve high throughput using conventional multi-access methods. However, when users are co-located and experience highly correlated channels, conventional multi-access schemes struggle to mitigate mutual interference, leading to significant performance degradation.
In such scenarios, users can instead exploit orthogonality in their semantic features via pre-trained semantic encoders. Moreover, semantic extraction  offers an additional degree of freedom. By adjusting the depth of semantic feature extraction, users can dynamically reshape their traffic demands to match the current transmission environment. This adaptive interplay between the channel and semantic domains motivates a unified framework that jointly optimizes channel control and semantic representation for multi-user communications.

While semantic communication presents a promising vision for future wireless networks, its practical deployment faces significant challenges. Existing semantic multi-access frameworks typically rely on strong assumptions or incur prohibitive training overhead. In particular, they often require a pre-trained semantic encoder-decoder for each transceiver pair~\cite{Zhang2023DeepMA,Li2023NOMA-Semantic}. This not only leads to substantial network-wide training costs but also severely compromises system flexibility and interoperability, rendering it infeasible between mismatched devices or newly joined users. Moreover, conventional semantic communication systems commonly assume fixed semantic compression ratios~\cite{Xu2022image_attention}. Such rigidity prevents dynamic adaptation to variations in semantic content, user priorities, and channel conditions, resulting in suboptimal spectral and energy efficiency in dynamic network environments.

Another challenge arises in optimizing transmission control parameters for multi-user semantic communications, where traffic demands, channel conditions, and semantic similarity are tightly coupled~\cite{Ma-twc2024}. Semantic encoder-decoder pairs are typically realized as large-scale deep neural networks (DNNs) with extensive parameter sets, whose performance is closely linked to user-specific channel characteristics and transmission control strategies. Generally, reliable training of semantic models relies on static or quasi-static channel conditions. However, when users update their transmission control variables, the effective channel conditions experienced by the semantic models change accordingly. This necessitates frequent model fine-tuning or full retraining, creating a feedback loop between transmission control and semantic model training. In dynamic and resource-constrained wireless networks, this interdependence becomes computationally prohibitive. Moreover, the integration of an RIS introduces high-dimensional discrete variables for phase-shift control while multi-user scheduling inherently involves combinatorial user grouping. These factors lead to a non-convex mixed-integer optimization problem that is intractable for conventional optimization techniques. Although deep reinforcement learning (DRL) presents a promising alternative for such complex decision-making problems, its direct application is often hindered by an explosively large state-action space and prohibitively high training overhead. These limitations motivate us to develop a more efficient optimization framework specifically designed for RIS-assisted multi-user semantic communication systems, which forms the focus of this work.

\subsection{Solutions and Contributions}
To address the aforementioned challenges, we first propose a joint source and RIS-assisted channel encoding (JSRE) framework for multi-user semantic communications. The objective is to maximize the system's aggregate energy efficiency by jointly optimizing the RIS's phase shifts, the semantic compression ratio, and the user scheduling strategy. One of our main contributions is the design of a single universal semantic encoder-decoder structure shared among all users, which significantly reduces model complexity and training overhead. Multi-user multiplexing and successful decoding are enabled by embedding user-specific channel state information (CSI) into the semantic encoding process. By leveraging the RIS's flexible channel reconfigurability, this universal semantic model achieves sufficient channel diversity to support concurrent multi-user transmissions while preserving high semantic similarity at each receiver. Moreover, the RIS's multi-band phase-shift control is exploited to enable differentiated channel enhancement across sub-bands. This allows on-demand traffic allocation over multiple sub-bands, thereby improving both spectral efficiency and end-to-end semantic fidelity under limited bandwidth constraints. Furthermore, the semantic encoder is designed to dynamically adjust its compression ratio based on semantic content and real-time channel conditions. Note that a lower compression ratio reduces computational load at the decoder and enhances robustness against interference, but it also increases traffic demands and imposes stricter requirements on resource allocation. Conversely, aggressive compression reduces data size at the cost of semantic accuracy and resilience. Motivated by this coupling analysis, we jointly optimize the RIS's phase shifts to shape the wireless channel and the semantic processing module to improve semantic accuracy, enabling adaptation to dynamic network conditions.

To jointly optimize the control parameters within the proposed JSRE framework, we propose a truncated DRL (T-DRL) framework that incorporates two special designs to alleviate the computational burden of the JSRE framework. First, we introduce a data-driven surrogate reward function, implemented as a dedicated DNN, which rapidly approximates system performance for any candidate control policy. Thus, it eliminates the need for costly environment rollouts during training. Second, to avoid mdoel retraining overhead, we design a semantic model caching mechanism that maintains a pool of pre-trained JSRE models corresponding to different user scheduling strategies. When the DRL agent generates a user scheduling strategy matching a cached entry, the corresponding JSRE model is retrieved directly, avoiding redundant retraining and thus significantly accelerating convergence. Specifically, the main contributions of this paper are summarized as follows:
\begin{itemize}
\item Unified semantic encoder-decoder model: The JSRE framework employs a shared semantic encoder-decoder model among all users. To enable concurrent multi-user transmissions, user-specific CSI is embedded directly into the semantic encoding process. This design jointly exploits channel orthogonality (induced by RIS's reconfiguration) and semantic orthogonality (inherent in users' semantic representations) to mitigate inter-user interference. We further propose a lightweight fine-tuning scheme for this shared architecture, allowing dynamic adjustment of the semantic compression ratio in response to real-time variations in semantic content and channel conditions. This capability significantly enhances system adaptability without requiring full model retraining.
\item Multi-band RIS control and traffic allocation: The multi-band RIS operates across multiple frequency sub-bands and achieves differentiated channel enhancement through correlated phase-shift control in each band. This capability enables the semantic traffic allocation among different sub-bands, i.e., users' semantic data can be dynamically distributed across sub-bands according to real-time interference levels and channel conditions. By aligning traffic demands with favorable spectral resources, the system improves both spectral efficiency and semantic fidelity under limited bandwidth.
\item T-DRL algorithm for fast learning in JSRE: We formulate an energy efficiency maximization problem that jointly optimizes the RIS's phase shifts, the semantic compression ratio, and the user scheduling strategy. We propose the T-DRL framework to approximate the high-dimensional and mixed-integer control problem, where the DRL agent can dynamically adapt the size of action space based on the current network state. To reduce training overhead, a data-driven similarity estimator is devised to rapidly predict semantic similarity under varying input conditions. Additionally, a semantic model caching mechanism is employed to store pre-trained models indexed by distinct user scheduling strategies. These components significantly enhance learning efficiency and reduce computational overhead during the learning process.
\end{itemize}

The JSRE framework was preliminarily validated in our previous conference paper~\cite{Wang-ispa2025}. This work significantly extends the system to a multi-band generalization scenario. Furthermore, we develop novel lightweight and efficient algorithms designed to ensure robust adaptability in dynamic network environments.
The remainder of this paper is organized as follows. Section~\ref{sec:related_works} reviews related work. Section~\ref{sec:system_model} details the system model and Section~\ref{sec:semantic model design} describes the JSRE framework and its training methodology. Section~\ref{sec:problem} presents the problem formulation and the T-DRL algorithm to optimize the JSRE framework. Section~\ref{sec:results} provides numerical results and discussions, and Section~\ref{sec:conclusion} concludes this paper.

\section{Related Work}\label{sec:related_works}
% Our work introduces a paradigm for wireless communication by synergistically integrating Reconfigurable Intelligent Surfaces (RIS), computational imaging principles, and the intrinsic sensitivity of deep neural networks. Rather than treating the RIS-sculpted wireless channel as a passive conduit to be optimized, we conceptualize it as an active computational component that—when coupled with a neural decoder—enables location-dependent information demultiplexing. This section outlines the three foundational pillars underpinning our approach, i.e., the evolution of RIS for spatially diverse channel engineering, advances in deep learning for semantic and multi-user communications, and the emerging notion of leveraging the physical layer for information encoding, inspired by adversarial machine learning.

\subsection{RIS-enhanced Channel Orthogonality}
An RIS enables precise control over the phase and amplitude of incident electromagnetic waves, allowing dynamic reconfiguration of the wireless propagation environment~\cite{Yashvanth2025IRSBeam}. The authors in~\cite{Pang2025IRS-joint-beamforming} focused on using an RIS to boost signal-to-noise ratio (SNR) and overall channel gain in single-user transmission scenarios. More recently, RIS has been extended to multi-user scenarios. Several studies integrated the RIS to suppress inter-user interference and improve both energy efficiency and spectral capacity~\cite{Sherif2024IRS-NOMA, Hao2025IRS_SDR1}. In our work, the research focus shifts from simple signal enhancement to enhancing spatial orthogonality. Thanks to its high spatial resolution, an RIS can generate distinct and differentiable channel responses for users at different physical locations. This capability has been mainly explored in physical layer security. By tailoring RIS's configurations, the authors in~\cite{Zhang2022PHY_DL_key} maximized channel orthogonality between legitimate users and eavesdroppers to increase secret key capacity. Building on this work, the authors in~\cite{Zhang2025RISSecure} further demonstrated the creation of spatially confined communication zones in multi-user settings. In these zones, only receivers located in designated regions can successfully decode the transmitted signal. These advances show that RIS can act not just as a channel booster, but also as a programmable filter that shapes the channel orthogonality.

\subsection{Deep Learning for Multi-User Semantic Communications}
Deep learning (DL) has profoundly reshaped physical-layer design. Joint source-channel coding (JSCC) frameworks now consistently outperform conventional approaches, especially in low-latency regimes~\cite{Jiang_Kim_Asnani_Kannan_Oh_Viswanath_2019}. This shift has catalyzed the rise of semantic communication, where the goal is no longer bit-perfect reconstruction but the reliable transmission of meaning. Extending semantic communication to multi-user settings remains an active research frontier. For example, the authors in~\cite{Li2023NOMA-Semantic} superimposed semantic symbols within a NOMA architecture, which shows performance gains over traditional bit-level NOMA. The authors in~\cite{Zhang2023DeepMA} proposed a DL-based multiple access (DeepMA) framework that jointly encodes and transmits semantic features from multiple users. However, DeepMA assigns a dedicated encoder–decoder pair to each user, leading to high training and storage overhead that limits its scalability. This drawback motivates our research in this work focusing on the design of universal semantic communication architectures. The authors in~\cite{Nash2025AirQuality} explored attention-based models that serve multiple users by separating their streams in latent space. In contrast, our approach rethinks this paradigm by jointly exploiting channel and semantic features to design a unified encoder-decoder architecture for all users.
\subsection{RIS-assisted Semantic Multiplexing}
Conventional multiple access schemes, such as time-, frequency-, and code-division, achieve a trade-off between spectral efficiency and transmission performance by linearly superimposing multiple coded streams~\cite{Li2025supercode}. As networks scale, wireless resources become insufficient to support a large number of users. To address this, recent work has explored semantic-level solutions by extracting semantic features at the source to improve multi-user access efficiency. The authors in~\cite{Ian2015Explain} showed that DNNs are highly sensitive to small perturbations in latent space. This observation motivates the use of DL methods to discover semantic orthogonality for multiple access. Moreover, semantic extraction from source data can adapt user traffic demands to dynamic channel conditions, which motivates the use of RIS to further manipulate channel orthogonality for semantic-level transmissions~\cite{Sherif2024IRS-NOMA}. For example, the authors in~\cite{2024-tcomJiang} leveraged RIS to improve channel conditions tailored to the transmission needs of different semantic information. By jointly exploiting RIS's reconfigurability and semantic control, the system gains additional degrees of freedom and increases network capacity. In this paper, we further exploit RIS to create user-specific channel signatures and embed them as features into the unified semantic communication model for all users. This eliminates the need for user-specific encoder-decoder pairs, significantly reducing training and deployment overhead.

 \section{Multi-Band RIS-assisted Multi-User Semantic Communication System}\label{sec:system_model}

\begin{figure}
    \centering{\includegraphics[width=0.48\textwidth]{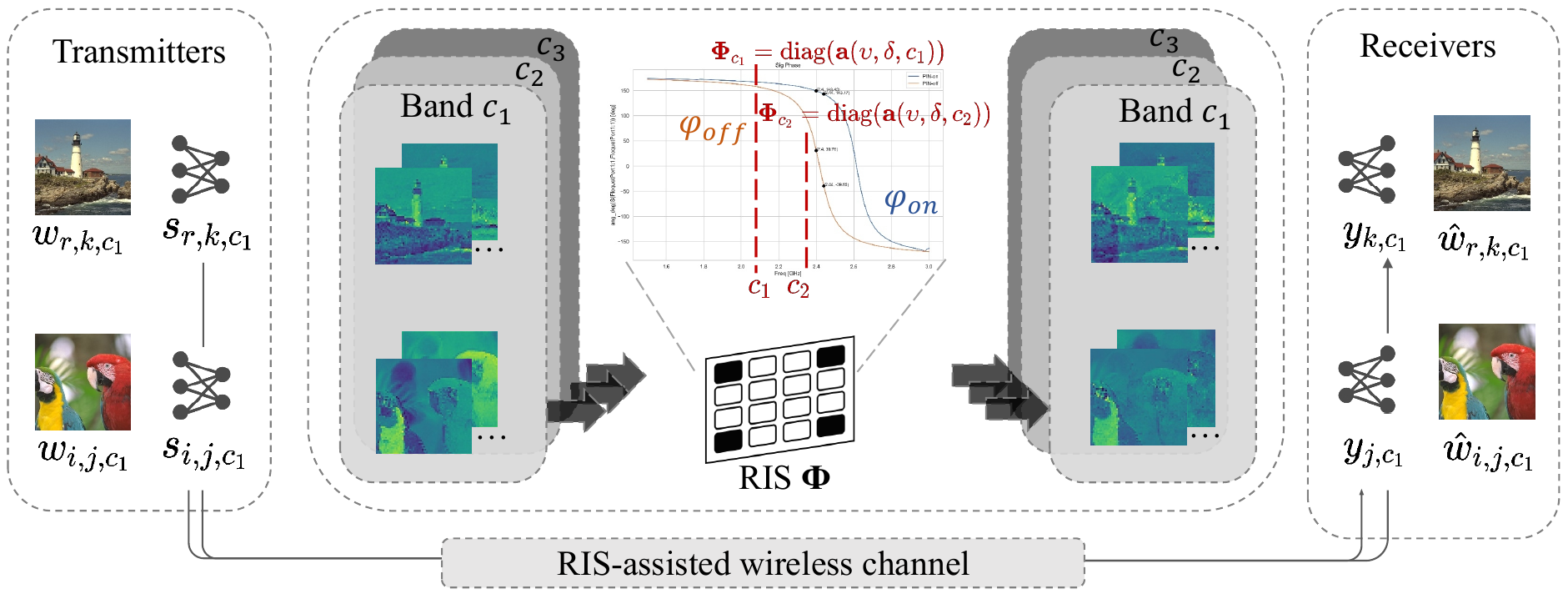}}
    % \vspace{-0.3cm}
    \caption{{The multi-band RIS-assisted multi-user semantic transmissions}.}
    \label{fig:system_model}
\end{figure}

As shown in Fig.~\ref{fig:system_model}, we consider an RIS-assisted multi-user semantic communication system. The set of users is denoted by $\mathcal{K} = \{1, \ldots, K\}$. We assume that each user pair has image or video transmission requirements to support diverse high-level tasks, such as urban traffic surveillance, disaster rescue operations, and environmental monitoring. Each user is equipped with a semantic encoder that extracts task-relevant semantic features from the raw data. The compressed semantic symbols are then transmitted over the wireless channel with reduced traffic data size and recovered at the corresponding receiver by a well-trained semantic decoder. To improve wireless channel conditions, a multi-band RIS is deployed on the ground to simultaneously configure the wireless channels in different bands. We assume that the system operates over $C$ orthogonal frequency bands, indexed by $c \in \{1, \ldots, C\}$. As such, each user can allocate its semantic traffic data over different bands, according to their channel conditions. We consider a long-term transmission scenario. For notational simplicity, we omit the time index in the following modeling.

\subsection{RIS-assisted Multi-band Channel Configuration}

Let $h_{k,r,c}$ denote the channel from user-$k$ to user-$r$ over band $c$ and the wireless channel is assumed to be reciprocal, i.e., $h_{k,r,c} = h_{r,k,c}$ for any user pair $(r, k) \in \mathcal{K} $. The channel from user-$k$ to the RIS over band $c$, denoted by $\mathbf{g}_{k,c}$, is assumed to follow Rician fading~\cite{Wu2025QoE_RIS} and given as follows:
\begin{equation}
    \mathbf{g}_{k,c} = \sqrt{\frac{\kappa}{\kappa+1}}\, \mathbf{g}^{\text{LoS}}_{k,c} + \sqrt{\frac{1}{\kappa+1}}\, \mathbf{g}^{\text{NLoS}}_{k,c},
    \label{equ:rician_channel}
\end{equation}
where $\kappa$ denotes the Rician factor quantifying the ratio of the line-of-sight (LoS) to the non-LoS (NLoS) components. The LoS component is computed as $\mathbf{g}^{\text{LoS}}_{k,c} = \alpha_d \mathbf{a}(\upsilon ,\delta, \bar{c})$, where the complex gain $\alpha_d$ represents path loss dependent on the distance $d_k$ between user-$k$ and the RIS. The coefficient $\mathbf{a}(\upsilon ,\delta, \bar{c})$ denotes the RIS's array response vector corresponding to the azimuth and elevation angles of arrival $(\upsilon ,\delta)$, given as follows:
\begin{equation}
    \begin{aligned}
        \mathbf{a}(\upsilon ,\delta, \bar{c}) = & \big[1,\dots,e^{j\frac{2\pi \bar{c}}{v}d_s(m\sin \upsilon \cos \delta + n\sin \delta)}, \\
        & \dots,e^{j\frac{2\pi \bar{c}}{v}d_s((P - 1)\sin \upsilon \cos \delta + (Q - 1)\sin \delta)}\big],
    \end{aligned}
\end{equation}
where $d_s$ represents the inter-element spacing and $\bar{c}$ is the speed of light. The RIS has $P$ elements along the horizontal dimension and $Q$ elements along the vertical dimension, resulting in a total of $N = PQ$ passive reflecting elements. The  NLoS component $\mathbf{g}^{\text{NLoS}}_{k,c}$ is modeled as an independent and identically distributed complex Gaussian random vector, i.e., $\mathbf{g}^{\text{NLoS}}_{k,c} \sim \mathcal{CN}\left( \mathbf{0}, \sigma^2 \mathbf{I}_N \right)$,  where $\sigma^2$ is the average power of the scattered components and $\mathbf{I}_N$ is the $ N \times N$ identity matrix.

For simplicity, we consider a $1$-bit control scheme for the RIS's phase shift strategy, where each element operates in one of two discrete states: ON or OFF state. Guided by empirical measurements~\cite{Scheder20241-bitIRS, Jana20242-bitIRS}, the frequency-dependent phase response of an RIS element on band $c$ can be approximated by a logistic-like function  as follows:
\begin{equation}\label{phase1}
    \phi_{\text{ON}}(c) = \frac{D_1}{1 + e^{a_1 c + b_1}} \text{ and }
    \phi_{\text{OFF}}(c)= \frac{D_2}{1 + e^{a_2 c + b_2}},
\end{equation}
where $a_1, a_2, b_1, b_2, D_1, D_2$ are device-specific constants determined by the RIS hardware characteristics. As shown in Fig.~\ref{fig:system_model}, the RIS's phase responses on different frequency bands are highly correlated. This implies that the performance enhancement provided by the RIS on one band is inherently coupled with that on the other bands. For example, a phase configuration that is favorable for band $c_1$ may inadvertently degrade the effective channel quality on another band $c_2$. Hence, the joint optimization of the RIS's phase shift and traffic allocation across multiple bands is essential to achieve overall system performance gains.

Let $\bm{\varphi}=[\varphi_1,\ldots,\varphi_N] \in \{0,1\}^N$, where ${\varphi}_n = 1$ activates the $n$-th RIS element in the ON state and ${\varphi}_n = 0$ sets it to the OFF state. As such, the complex reflection coefficient of the $n$-th RIS element on band $c$ is defined as follows:
\begin{equation}\label{phase2}
\psi_n(c) =
\begin{cases}
e^{j\phi_{\text{ON}}(c)}, & \text{if } {\varphi}_n = 1, \\
e^{j\phi_{\text{OFF}}(c)}, & \text{if } {\varphi}_n = 0.
\end{cases}
\end{equation}
We consider a challenging propagation environment in which direct user-to-user links are severely attenuated due to blockages and thus can be neglected~\cite{Guo2023block}. Thus, the effective channel from user-$r$ to user-$k$ is represented as $h_{r,k,c} = \mathbf{g}^H_{r,c} \boldsymbol{\Phi}_c \mathbf{g}_{k,c}$, where $\boldsymbol{\Phi}_{c} = \mathrm{diag}\big(\psi_1(c), \psi_2(c), \dots, \psi_N(c)\big) \in \mathbb{C}^{N \times N}$ is the diagonal reflection matrix of the RIS on band $c$.

{
\iffalse
\begin{figure}
    \centering
    \includegraphics[width=0.48\textwidth]{figures/multi-band_phi.pdf}
    \caption{The joint control of multi-band reflecting coefficients}
    \label{fig:multi_band_phi}
\end{figure}
\fi
}

\subsection{Semantic Extraction and Recovery}

\iffalse
1) explain semantic extraction at tx and recovery at rx, especially how to obtain variable length semantic information $s_{s,k}\in\mathcal{R}^L$ with length $L$
\[
s_{raw} \in\mathcal{R}^M  --> s_{s,k} \in\mathcal{R}^L
\]
explain how $L$ is related to compression ratio.  explain the DNN model for semantic extraction. A larger $L$... A smaller $L$ ...
Let $s_{r,k}^w \in\mathcal{R}^M $ denote the raw data.
\[
s_{r,k} = \text{semext}(s_{r,k}^w, \theta) \in\mathcal{R}^L
\]
where function $\text{semext}(\cdot)$ denotes the semantic extraction from the raw data $s_{r,k}^w$. $\theta$ denotes the semantic compression ratio.
\fi

Each user can dynamically adjust its data traffic by extracting task-relevant semantic information from the raw data.
Let $w_{r,k,c}\in\mathbb{R}^M$ denote the raw data of dimension $M$ transmitted between user pair $(r,k)$ over band $c$. Through semantic extraction, it can be mapped to a compact representation  $s_{r,k,c} \in \mathbb{R}^L$, where $L$ reflects the reduced size of traffic demand, i.e.,~$L \ll M$. Let $\beta_{r,k}$ denote the semantic compression ratio during the semantic transmission from user-$r$ to user-$k$. To enable adaptive and channel-aware semantic encoding, we employ a unified DNN-based semantic encoder $\mathcal{E}_{\theta}$, parameterized by a shared set of learnable weights $\theta$, whose input consists of the user's raw data $w_{r,k,c}$, the user-specific channel $h_{r,k,c}$, and the semantic compression ratio $\beta_{r,k}$:
\begin{equation}
    s_{r,k,c} = \mathcal{E}_{\theta}\bigl(w_{r,k,c},\, h_{r,k,c}, \beta_{r,k}\bigr).
\end{equation}

Let $b_{r,k,c}$ denote the binary scheduling decision on frequency band $c$. When $b_{r,k,c}=1$, it indicates that user-$r$ is scheduled to transmit its semantic information to user-$k$, while $b_{r,k,c}=0$ implies that user-$r$ remains silent on this channel. We define the user scheduling strategy as $\mathbf{B}=\{b_{r,k,c}\}_{r,k\in\mathcal{K},c\in\mathcal{C}}$. Hence, the received signals at user-$k$ on band $c$ is expressed as follows:
\begin{equation}
    y_{k,c} = \sum_{r\in\mathcal{K}, r\neq k } b_{r,k,c} h_{r,k,c} s_{r,k,c} + n_{k,c},
    \label{equ:propagation process}
\end{equation}
where $n_{k,c}$ represents the noise at user-$k$ on band $c$. Note that $\mathbf{B}$ determines the interference level in each semantic transmission. At the receiver side, each user employs a DNN-based semantic decoder $\mathcal{E}_{\tilde{\theta}}^{-1}$, parameterized by $\tilde{\theta}$, to reconstruct the original raw data from the received signal $y_{k,c}$. Thus, the recovered data $\hat{w}_{r,k,c}$ is represented as follows:
\begin{equation}
\hat{w}_{r,k,c} = \mathcal{E}_{\tilde{\theta}}^{-1}\left(y_{k,c}, \, h_{r,k,c}, \beta_{r,k} \right).
\label{equ:decode}
\end{equation}
With well-trained encoder-decoder parameters $({\theta},\tilde{\theta})$, each user is expected to accurately reconstruct the raw data ${w}_{r,k,c}$.

%However, semantic recovery may practically fail due to channel fluctuations and improper user scheduling that bring eccessive interference in semantic decoding.

To evaluate the performance of semantic recovery, we adopt the structure similarity (SSIM) metric $\xi_{r,k,c}$ to characterize the similarity between $w_{r,k,c}$ and the recovered data $\hat{w}_{r,k,c}$~\cite{Yan2022SemRate4text}:
\begin{equation}\label{equ:ssim}
    \xi_{r,k,c} = \frac{(2\mu_{w_{r,k,c}}\mu_{\hat{w}_{r,k,c}}+\zeta_1)(2\sigma_{w_{r,k,c}\hat{w}_{r,k,c}}+\zeta_2)}{(\mu^2_{w_{r,k,c}}+\mu^2_{\hat{w}_{r,k,c}}+\zeta_1)(\sigma^2_{w_{r,k,c}}+\sigma^2_{\hat{w}_{r,k,c}}+\zeta_2)},
\end{equation}
where $\mu_{x}$ and $\sigma_{x}$ are the sample mean value and standard derivation of data $x$, respectively. The coefficient $\sigma_{w_{r,k,c}\hat{w}_{r,k,c}}$ denotes the sample  covariance between $w_{r,k,c}$ and $\hat{w}_{r,k,c}$. The constants $\zeta_1$ and $\zeta_2$ are introduced to avoid numerical instability when the denominator approaches zero. Given the SSIM-based fidelity measure, we define the semantic rate $\Gamma_{r,k,c}$ from user-$r$ to user-$k$ on band $c$ as follows:
\begin{equation}
    \Gamma_{r,k,c} = \frac{W_c S_{\text{sem}} \xi_{r,k,c} }{\beta_{r,k} M},
    \label{equ:gamma}
\end{equation}
where $W_c$ is the bandwidth of band $c$ and $S_{\text{sem}}$ denotes the average amount of semantic information carried in raw data. The numerator in~\eqref{equ:gamma} represents the amount of successfully delivered semantic information, while the denominator represents the cost of transmitted symbols. Thus, semantic rate $\Gamma_{r,k,c}$ provides a measure of semantic transmission efficiency that jointly accounts for the communication resources, the semantic compression ratio, and the recovery quality.

\subsection{Energy Consumption and System Efficiency}
Let $F(\beta_{r,k})$ denote the number of floating-point operations (FLOPs) required for semantic encoding and decoding of the user pair $(r,k)$  with the compression ratio $\beta_{r,k}$. The semantic processing latency can be evaluated as $t^s_{r,k} = F(\beta_{r,k}) / (n_G f_G)$, where $f_G$ denotes the GPU clock frequency and $n_G$ denotes the number of FLOPs that is executed per clock cycle~\cite{Ji2024SemanticEE}. Thus, the energy consumption is computed as $p_s t^s_{r,k}$, where $p_s$ denotes the constant power consumption in semantic feature extraction and recovery. The total semantic processing energy for all user pairs can be evaluated as follows:
\begin{equation}
    E_s =  \sum_{r, k \in\mathcal{K}, c\in\mathcal{C}} b_{r,k,c} p_s t^s_{r,k}.
\end{equation}
After semantic extraction, each semantic symbol $s_{r,k,c}$ is transmitted with an average power $p_{d}$. Let $t^{d}_{r,k,c} = L / W_c$ denote the data delivery time over band $c$. Hence, the  total transmission energy consumption is represented as follows:
\begin{equation}
    E_t = \sum_{r, k \in\mathcal{K}, c\in\mathcal{C}} b_{r,k,c} p_d t^{d}_{r,k,c}.
\end{equation}
The RIS also incurs a constant energy consumption $E_r = N e_r $, where $e_r$ is the per-element energy consumption~\cite{Ye2025IRS_Engergy}. Hence, the overall energy consumption is $E_s + E_t + E_r$. Accordingly, the system's energy efficiency $\eta$ is defined as the ratio of the aggregate semantic rate to the total energy consumption:
\begin{equation}
    \eta = \frac{\sum_{r, k \in\mathcal{K}, c\in\mathcal{C}} b_{r,k,c} \Gamma_{r,k,c}}{E_s + E_t + E_r}.
    \label{equ:EE}
\end{equation}
The energy efficiency $\eta$ is jointly determined by user scheduling strategy, the RIS's phase shifts, and the semantic compression ratio. Obviously, transmitting a larger amount of semantic information improves semantic recovery accuracy but incurs larger transmission delay and energy consumption. Conversely, aggressive semantic compression reduces traffic demand and transmission overhead, with the cost of reduced semantic recovery quality. Therefore, it is crucial to balance the trade-off between semantic fidelity and transmission efficiency.

\begin{figure*}
    \centering
    \includegraphics[width=0.99\textwidth]{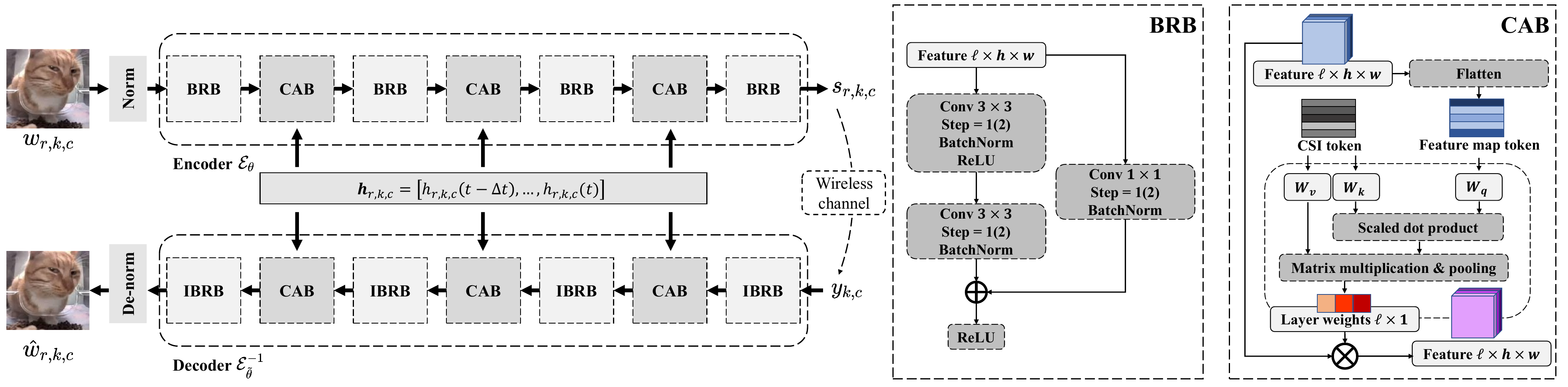}
    \caption{The JSRE's network architecture for semantic encoder and decoder shared by all users.}
    \label{fig:network_architecture}
    \vspace{-0.3cm}
\end{figure*}

\section{Network Architecture for JSRE Framework}~\label{sec:semantic model design}

In dynamic wireless environments, CSI varies over time. This variation poses a significant challenge to conventional semantic encoder-decoder models that are typically trained under a fixed CSI condition. To address this issue, we propose the JSRE framework, which explicitly incorporates varying CSI into the semantic encoding and decoding process. By treating user-specific CSI as an input feature, the JSRE framework adapts to time-varying channels and supports reliable semantic transmissions under diverse propagation conditions. Another key advantage of incorporating CSI in the JSRE framework is that it enables a unified semantic encoder-decoder model for all users. Such user-specific CSI information can help differentiate users even when they transmit individual semantic feature symbols simultaneously via a shared encoder-decoder model.

As illustrated in Fig.~\ref{fig:network_architecture}, the semantic encoder consists of multiple basic residual blocks (BRBs) that progressively extract semantic features from the raw data. Channel attention blocks (CABs) are embedded within the encoder to fuse CSI with intermediate features, allowing the model to adapt its semantic feature extraction based on current channel conditions. At the receiver, the semantic decoder adopts a similar structure, replacing convolutional layers with deconvolutional layers and using inverse basic residual blocks (IBRBs) to recover the original semantic content. The proposed JSRE framework offers two main advantages. First, the semantic compression ratio can be dynamically adjusted by varying the depth of the BRB layers in the encoder. This flexibility allows a joint optimization of semantic feature extraction and RIS-assisted transmission control, increasing the degrees of freedom in end-to-end semantic communication. Second, by sharing a single encoder-decoder model among all users and leveraging RIS's phase shifts to manage user-specific channel characteristics, this framework eliminates the need for per-user or per-link model training. This greatly reduces training overhead and improves scalability to large-scale wireless networks.

\subsection{Flexible Semantic Compression via BRB Modules}

The semantic encoder stacks BRBs and CABs alternately to fuse semantic features with user-specific CSI, following a similar design principle in~\cite{Xu2022image_attention} and~\cite{Zhang2023DeepMA}.  By extracting the semantic representation from different stages of the BRB module, the system can achieve varying levels of semantic compression ratios, as shown in Fig.~\ref{fig:network_architecture}. In our implementation, each BRB module comprises two  $3\times3$ convolutional layers with padding of $1$, preserving spatial resolution within the block. A skip connection with a $1\times1$ convolution is incorporated to align the dimensionality between the input and output feature maps. Combined with the ReLU activation, this residual structure stabilizes training and alleviates gradient vanishing.

Let $\ell$ denote the number of layers of the input feature map, and $(h,w)$ denote the height and width in pixels in each feature layer. Given the BRB network structure, the first BRB module downsamples the input feature map from $(\ell, h, w)$ to $(3\ell, h/2, w/2)$. Each subsequent BRB applies a similar spatial downsampling while expanding the dimension of feature layers. As such, the input feature map can be compressed by $3 \over 4$ after each BRB. Hence, the output from the $i$-th BRB block offers a compression ratio of $\beta_{r,k}=(\frac{3}{4})^i$. This process continues by enlarging the feature layers while compressing the height and width until we reach a desired compression ratio $\beta_{r,k}$ for semantic transmission. Finer-grained compression ratios can be also achieved by properly designing the size of convolutional layers in different BRB modules.

\subsection{CSI Integration via CAB Modules}
The CAB modules are employed to incorporate user-specific channel features into the semantic representations. Each CAB connects two consecutive BRBs by first tokenizing the intermediate feature map and then concatenating it with the corresponding channel feature vector. To capture temporal channel dynamics, we introduce a measurement window of length $\Delta t$ that records historical channel observations for each user. Specifically, the CSI for user pair $(r,k)$ on band $c$ at time $t$ is recorded as $ {\bm h}_{r,k,c}=[h_{r,k,c}(t-\Delta t), \dots, h_{r,k,c}(t)]$. This channel vector is embedded and integrated into the convolutional semantic encoder through a cross-attention mechanism~\cite{Xu2022image_attention}. We adopt channel-wise attention rather than spatial attention to avoid introducing interference in the spatial domain.

Each input feature map $s_{r,k,c}$ in the CAB is first flattened along the spatial dimensions with a unified embedding dimension $d$, and then concatenated along with the layer dimension, i.e., $ \text{vec}(s_{r,k,c}) \in \mathbb{R}^{\ell\times d}$.
The cross-attention mechanism employs three learnable projection matrices, denoted as $\mathbf{W}_q \in \mathbb{R}^{d \times d}$, $\mathbf{W}_k \in \mathbb{R}^{d \times d}$, and $\mathbf{W}_v \in \mathbb{R}^{d \times d}$, to capture the latent dependencies between the semantic features and the channel conditions~\cite{Dong-tsp2025}. Accordingly, the  resulting channel attention $a_{r,k,c}$ is computed as follows:
\begin{subequations}\label{equ:attention}
\begin{align}
    & \mathbf{Q}_{r,k,c}=\text{vec}(s_{r,k,c})\mathbf{W}_q, \\
    & \mathbf{K}_{r,k,c}=\bm{h}_{r,k,c}\mathbf{W}_k, \\
    & \mathbf{V}_{r,k,c}=\bm{h}_{r,k,c}\mathbf{W}_v, \\
    & a_{r,k,c} = \text{softmax}\left(\frac{\mathbf{Q}_{r,k,c}\mathbf{K}_{r,k,c}^T}{\sqrt{d}}\right)\mathbf{V}_{r,k,c},
\end{align}
\end{subequations}
where $\mathbf{Q}_{r,k,c}$, $\mathbf{K}_{r,k,c}$, and $\mathbf{V}_{r,k,c}$ denote the query, key, and value matrices, respectively. The attention output $a_{r,k,c} \in \mathbb{R}^{\ell\times d}$ has the same dimensions as the flattened input feature $\text{vec}(s_{r,k,c})$. This output encodes the channel-conditioned relevance between each semantic token and the instantaneous CSI. Unlike standard transformers~\cite{vaswani2023attention}, where attention outputs are incorporated through additive feature refinement, our goal is to selectively retain the most suitable feature layers conditioned on CSI. Therefore, we apply pooling and normalization to $a_{r,k,c}$ to derive a gated weight vector, which adaptively modulates the feature layers during feature encoding.

\subsection{Training with Hierarchical Perceptual Loss}
To enable high-fidelity image reconstruction  from semantic features extracted at arbitrary layers of the encoder, the overall semantic loss function $\mathcal{L}_{\text{sem}}$ integrates two complementary objectives. The first objective ensures accurate end-to-end image reconstruction, while the second enforces hierarchical consistency of semantic feature representations between corresponding layers of the encoder and decoder. First, to ensure pixel-level fidelity, we adopt the mean squared error (MSE) as the primary reconstruction loss $\mathcal{L}_{\text{recon}} = \frac{1}{I} \sum_{i=1}^{I} \|x_i - \hat{x}_i\|^2$, where $I$ denotes the total number of pixels, and $x_i$ and $\hat{x}_i$ represent the ground-truth and reconstructed pixel values of the image data, respectively.
Second, inspired by perceptual loss formulations~\cite{johnson2016perceptuallossesrealtimestyle}, we introduce a hierarchical feature consistency loss to enforce semantic alignment across different network depths. Specifically, let $f_e^j$ and $f_d^j$ denote the intermediate feature maps at the $j$-th depth of the encoder and decoder, respectively. The feature consistency loss is defined as $\mathcal{L}_{\text{feat}} = \sum_{j=1}^{D} \|f_e^j - f_d^j\|^2$, where $D$ is the number of aligned depth levels in the encoder-decoder structure. This term encourages the decoder to preserve not only global image structure but also the multi-scale semantic content embedded in intermediate representations. Thus, the overall semantic loss is defined as a weighted combination of the reconstruction and feature consistency losses as follows:
\begin{equation}
    \mathcal{L}_{\text{sem}} = \lambda \, \mathcal{L}_{\text{recon}} + (1 - \lambda) \, \mathcal{L}_{\text{feat}},
\end{equation}
where $\lambda \in [0,1]$ is a tunable hyperparameter that balances pixel-wise accuracy against hierarchical semantic consistency.  By minimizing $\mathcal{L}_{\text{sem}}$, the encoder-decoder model learns a compressed semantic representation that supports both high-fidelity image recovery and robust multi-scale feature alignment.

\section{Maximizing JSRE's Energy Efficiency}\label{sec:problem}
The JSRE framework provides a unified encoder-decoder model tailored for multi-user semantic communications. Note that there exists an inherent trade-off between communication efficiency and semantic fidelity. A higher compression ratio $\beta_{r,k}$ reduces the traffic demand for user pair $(r,k)$ but degrades the reconstruction accuracy at the receiver, whereas transmitting less compressed semantic features improves semantic fidelity at the cost of increased transmission load. To maximize the overall energy efficiency of the proposed JSRE framework, we formulate a joint optimization problem of the semantic compression ratio $\bm\beta \triangleq \{\beta_{r,k}\}_{r,k\in\mathcal{K}}$, the user scheduling ${\bf B}=\{b_{r,k,c}\}_{r,k\in\mathcal{K},c\in\mathcal{C}}$, and the RIS's phase shift $\boldsymbol{\Phi}=\{\boldsymbol{\Phi}_{c}\}_{c\in\mathcal{C}}$ as follows:
\begin{subequations}\label{equ:energy_efficiency_problem}
\begin{align}
\max_{\bf{B}, \boldsymbol{\Phi}, \bm\beta} ~&~ {  \mathbb{E}[\eta]} \label{prob-ee-obj} \\
\mathrm{s.t.}
    ~&~\mathbb{E}\left[ \sum_{k\in\mathcal{K},c\in\mathcal{C}}\Gamma_{r,k,c} \right] \geq \Gamma_{\min},  \label{main_problem_constraint:fair}\\
    ~&~\sum_{k\in\mathcal{K}}b_{r,k,c}+b_{k,r,c}\leq 1, \label{main_problem_constraint:half_duplex}\\
    ~&~\beta_{r,k}\in (0,1], \eqref{phase1}  \text{ and } \eqref{phase2}, \label{main_problem_constraint:RIS}\\
    ~&~ r\in\mathcal{K}, k\in\mathcal{K}, \text{ and } c\in\mathcal{C},
\end{align}
\end{subequations}
where the expectation $\mathbb{E}[\cdot]$ in~\eqref{prob-ee-obj} is taken over the long-term dynamics of channel conditions and users' traffic demands. Constraint~\eqref{main_problem_constraint:fair} ensures that each user achieves a minimum expected semantic rate $\Gamma_{\min}$, thus guaranteeing baseline quality of service. Constraint~\eqref{main_problem_constraint:half_duplex} enforces half-duplex operation by restricting each user to at most one active transmission link per band. Problem~\eqref{equ:energy_efficiency_problem} is challenging to solve directly for two main reasons. First, the energy efficiency metric $\eta$ depends on the numeric SSIM metric, which lacks a closed-form analytical expression. Second, the problem is a mixed-integer nonconvex optimization task. These characteristics render conventional model-based optimization methods computationally intractable or impractical in dynamic environments. To address these challenges, we adopt a model-free DRL approach by learning near-optimal policies through repeated interactions with the environment, without requiring explicit knowledge of the underlying system dynamics. In the sequel, we detail the proposed DRL algorithm, focusing on two specially designed architectural enhancements to improve the learning performance in the proposed JSRE framework.

\subsection{DRL for Energy Efficiency Maximization}
We reformulate the optimization problem~\eqref{equ:energy_efficiency_problem} as a Markov decision process (MDP) with a tuple ($\mathcal{O}, \mathcal{A},\mathcal{R}$), where $\mathcal{O}$, $\mathcal{A}$, and $\mathcal{R}$ denote the state, action, and reward spaces, respectively. To capture the users' channel conditions and traffic demands, we define the state as ${\bm o} = \{ {\bm h}, \boldsymbol{\Phi}, {\bm n} \}\in\mathcal{O}$ at each time slot, where ${\bm h}=\{h_{r,k,c}\}_{r,k\in\mathcal{K},c\in\mathcal{C}}$.  We also maintain a counter vector ${\bm n}  = \{{ n}_{1} , { n}_{2} , \dots, { n}_{K} \} $, where ${n}_{k} $ denotes the user-$k$'s cumulative transmission times up to the current time slot. This state representation enables the agent to balance real-time channel quality and long-term fairness through awareness of historical resource allocation. The action ${\bm a} = \{\bf{B}, \boldsymbol{\Phi}, \bm\beta\}\in \mathcal{A}$  consists of the user scheduling matrix, the RIS's phase shifts, and the semantic compression ratio. The instantaneous reward $r \in\mathcal{R}$ evaluates the effectiveness of the state-action pair adopted in each decision round and guides the users toward long-term system performance improvement.
To this end, the reward function $r$ can be intuitively formulated as follows:
\begin{equation}\label{equ-reward}
r = \eta - \vartheta \sum_{r \in \mathcal{K}} \left( \Gamma_{\min} - \mathbb{E}_{\Delta t} \left[ \sum_{k \in \mathcal{K},c \in \mathcal{C}} \Gamma_{r,k,c} \right] \right)^+,
\end{equation}
where $\mathbb{E}_{\Delta t}[\cdot]$ denotes the empirical average over a sliding time window of duration $\Delta t$ and $(x)^+ \triangleq  \max\{0,x\}$.  The coefficient $\vartheta > 0$ is a large penalty parameter. 

We adopt the proximal policy optimization (PPO) algorithm to solve problem~\eqref{equ:energy_efficiency_problem}~\cite{Zhang-2024jsac}, which maintains the actor network to learn a stochastic policy $\pi_{\theta_a}(\bm{a}|\bm{o})$ and the critic network to estimate the state-value function $V_{\theta_c}(\bm{o})$, parameterized by $\theta_a$ and $\theta_c$, respectively. To improve stable learning, PPO introduces a clipped surrogate objective to limit the step size for the policy update. Let $\rho_t  =  \pi_{\theta_a} / \pi_{\theta_{\text{old}}} $ denote the probability ratio between the new policy $\pi_{\theta_a}$ and old policy $\pi_{\theta_{\text{old}}}$. The policy network is updated to improve the expected clipped surrogate function as follows:
\begin{equation}
    J(\theta_a) = \mathbb{E}_{t} \left[ \min \left( \rho_t  \hat{A},\; \text{clip}\bigl(\rho_t , 1-\epsilon, 1+\epsilon\bigr) \hat{A} \right) \right],
\end{equation}
where $\hat{A}$ represents the advantage value estimator and $\epsilon$ is a hyperparameter determining the clipping range. The function $\text{clip}(\cdot)$ restricts the probability ratio $\rho_t$ within the safe region $[1-\epsilon, 1+\epsilon]$, preventing the new policy from deviating excessively from the behavior policy. The critic network parameter $\theta_c$ is updated by minimizing the MSE loss as follows:
\begin{equation}
    L(\theta_c) = \mathbb{E}_{t} \left[ \bigl(V_{\theta_c}(\bm{o}) - V^{\text{targ}}\bigr)^2 \right],
\end{equation}
where $V^{\text{targ}}$ is the target value generated from a delayed critic network. Finally, the complete learning objective combines the policy improvement, value function error, and an entropy bonus to encourage exploration as follows:
\begin{equation}
    \max_{\theta_a, \theta_c} \tilde J(\theta_a, \theta_c) = \mathbb{E}_{t} \left[ J(\theta_a) - \eta_1 L(\theta_c) + \eta_2 S({\theta_a}) \right],
    \label{equ:PPO_objective}
\end{equation}
where $\eta_1$ and $\eta_2$ balance the trade-off between the value loss and the entropy bonus $S(\theta_a)$. 

Although PPO can directly solve problem~\eqref{equ:energy_efficiency_problem}, its practical implementation faces two significant challenges. One challenge is that the semantic similarity measure is required in the reward function~\eqref{equ-reward}, which lacks an explicit analytical expression. Its numerical evaluation depends on interaction within the semantic transmission system to generate sufficient sample data. Another challenge arises from the high-dimensional and heterogeneous nature of the decision variables. These difficulties make it hard to guarantee the effectiveness of training process. To address these challenges, we propose two key enhancements to the model-free PPO algorithm. First, instead of evaluating semantic similarity through costly interactions with the environment, we train an auxiliary neural estimator to predict the SSIM metric in real time based on current system states. This surrogate model enables fast reward computation without requiring costly observations of the actual semantic transmission system during training. Second, we introduce a truncated action generation mechanism within the DRL architecture, which dynamically shapes the size of action search space according to the observed state. The truncation of the action space is realized through a model caching mechanism, which leverages pre-trained semantic models enhanced with low-rank adaptation. This approach avoids redundant retraining of high-dimensional DNN components, significantly reducing computational overhead and improving sample efficiency.
  %Building on these two innovations, we detail the  T-DRL framework as below.

\begin{figure}[t]
    \centering
    \includegraphics[width=0.48\textwidth]{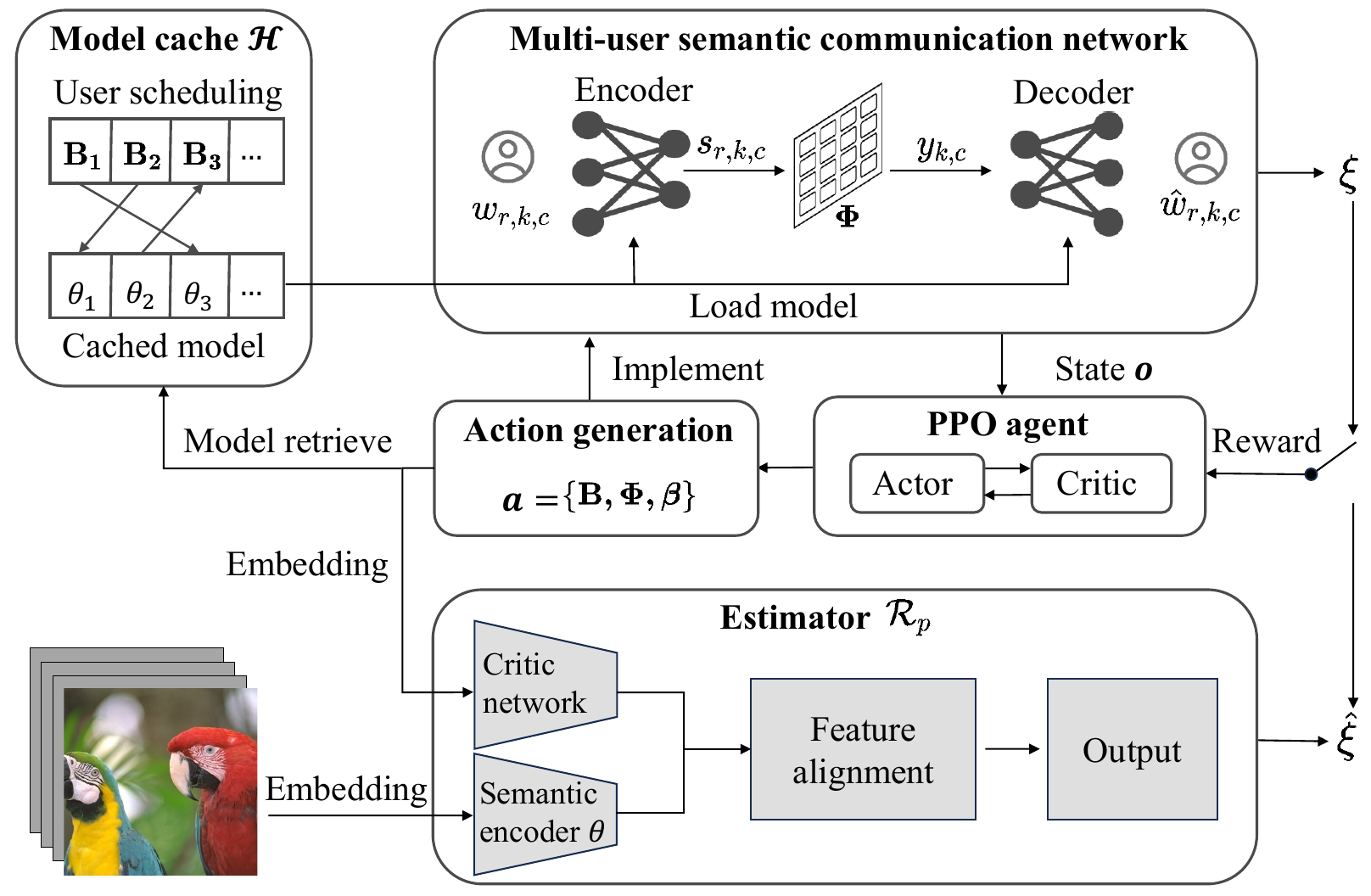}
    \caption{The T-DRL framework with fast reward estimator and model caching.}
    \label{fig:algorithm_framework}
\end{figure}
\subsection{Fast Learning with Semantic Similarity Estimator}
As defined in~\eqref{equ:ssim}, the semantic similarity is numerically evaluated as the similarity between the raw source image and the semantically reconstructed image data at the receiver. However, this direct evaluation incurs significant signaling and computational overhead during DRL training, as it requires performing extensive semantic transmissions for every interaction with the environment. To circumvent this bottleneck, we design a DNN-based estimator $\mathcal{R}_p$ that rapidly evaluates the semantic similarity $\xi_{r,k,c}$ for any user pair $(r,k)$ given the input raw image and the current control variables $(\bf{B}, \boldsymbol{\Phi}, \bm\beta)$. As shown in Fig.~\ref{fig:algorithm_framework}, the estimator $\mathcal{R}_p$ has two inputs: one processes the raw input image through a convolutional feature extractor and the other encodes the structured control variables $(\bf{B}, \boldsymbol{\Phi}, \bm\beta)$ using fully connected layers. After extracting feature representations from both branches, the two modalities are dimensionally aligned and fused. Then, the fused representation is passed through a fully connected prediction layer to produce the estimated semantic similarity score $\hat{\xi}_{r,k,c}$. The estimator's network structure is detailed as follows:

\subsubsection{Estimator structure and training scheme}
The estimator's image input branch directly adopts the encoder of the pre-trained semantic model, eliminating the need to train a separate vision backbone. As illustrated in Fig.~\ref{fig:algorithm_framework}, during data collection, the semantic transmission environment is configured according to a given set of control variables $(\bf{B}, \boldsymbol{\Phi}, \bm\beta)$, and the resulting semantic similarity $\xi_{r,k,c}$ is recorded as ground-truth supervision.
The embedding layers for the control variables are shared with the critic network in the DRL agent. This parameter sharing not only reduces redundancy but also promotes consistency between reward estimation and policy evaluation. Image features and control variable embeddings are first processed through two separate multilayer perceptrons (MLPs) to align their representations and reduce dimensionality. The resulting feature vectors are then concatenated and passed through a third MLP to produce a semantic similarity score. These three MLPs form the trainable components of the estimator. Notably, the embedding modules themselves are not jointly trained, which substantially reduces the number of trainable parameters. During training, the DRL agent first explores the environment by trying different control settings and recording the true semantic similarity. These experiences are stored in a replay buffer. The estimator learns to mimic this behavior by predicting the correct similarity from the same inputs, thus enabling fast reward evaluation without performing extensive semantic transmissions.

\subsubsection{Model-based warm start}
The amount of available training data is often very limited. To facilitate convergence under such data-scarce conditions, we adopt a model-based approach to expedite the estimator's training process. Observations from the state-of-the-art literature~\cite{Letafati2025JSCC4image,Li2023NOMA-Semantic,Yan2022SemRate4text} indicate that the semantic similarity $\xi_{r,k,c}$ is a monotonically increasing function of both the signal-to-interference-plus-noise ratio (SINR) $\gamma_{r,k,c}$ and the compression ratio $\beta_{r,k}$ for the user pair $(r,k)$'s semantic transmission. Such observations can be leveraged to initialize the reward estimator before fine-tuning it with the actual transmission data. Specifically, the SINR from user-$r$ to user-$k$ over band $c$ can be derived as follows:
\begin{equation}
    \gamma_{r,k,c} = \frac{p_t|h_{r,k,c}|^2}{\sigma^2+\sum^K_{\substack{i=1,i\neq {r}}}\sum^K_{\substack{j=1,j\neq {k}}}b_{i,j,c}p_t|h_{j,k,c}|^2},
    \label{equ:SINR}
\end{equation}
where $\sigma^2$ denotes the noise power. Based on the measurement results in~\cite{Zhang2023DeepMA,Letafati2025JSCC4image,Yan2022SemRate4text}, the relationship between $\xi_{r,k,c}$ and $(\gamma_{r,k,c}, \beta_{r,k})$ can be exponentially approximated as follows:
\begin{equation}
\xi_{r,k,c}=1-\exp\{-(k_1\gamma_{r,k,c}+b_1)(k_2\beta_{r,k}+b_2)\},
    \label{equ:surface}
\end{equation}
where the constant coefficients \(k_1\), \(b_1\), \(k_2\), and \(b_2\) are intrinsic properties of the semantic communication system and depend on channel conditions, encoding-decoding algorithms, and the transmitted data.
Therefore, given this approximation, we can employ~\eqref{equ:surface} to generate a large number of synthetic samples to initialize the estimator's model weights. Subsequently, the estimator is fine-tuned using actual experiences stored in the replay buffer. As such, given the control variables $(\mathbf{B}, \boldsymbol{\Phi}, \bm\beta)$, the estimator can predict the semantic similarity $\hat{\xi}_{r,k,c} = \mathcal{R}_p(\mathbf{B}, \boldsymbol{\Phi}, \bm\beta)$. During training, the mean squared error loss $\mathcal{L}_{\text{est}} = \mathbb{E} \left[ \lVert\xi_{r,k,c} - \hat{\xi}_{r,k,c}\rVert^2 \right]$ is employed. Once the training loss converges, the output $\hat{\xi}_{r,k,c}$ of the estimator replaces the true semantic similarity $\xi_{r,k,c}$ in the reward calculation, enabling fast evaluation during the DRL training process.

\subsection{Reducing Training Overhead via Model Caching}
The transmission control variables $(\mathbf{B}, \boldsymbol{\Phi}, \bm\beta)$ shape the transmission environment, necessitating continuous fine-tuning of the semantic model for effective adaptation. This incurs significant training overhead, especially in time-varying network environments. A key observation is that the user scheduling decision $\bf{B}$ is the most salient factor affecting the transmission environment. Any change in $\bf{B}$ alters the multipath channel conditions and interference levels, thereby requiring the semantic model to be fine-tuned accordingly. Conversely, when $\bf{B}$ remains fixed, the existing semantic encoding model can be reused, and further performance gains can still be achieved by tuning the other control variables, i.e., the semantic compression ratio $\bm\beta$ and the RIS's phase shifts $\boldsymbol{\Phi}$. This insight motivates a model caching strategy that maintains a one-to-one mapping from each user scheduling decision $\bf{B}$ to the semantic model parameters $\theta$. As such, we can avoid redundant retraining of the model parameters within DRL iterations when the user scheduling decision $\mathbf{B}$ is not changed, thereby substantially reducing computational overhead.
\subsubsection{Model caching scheme}
As illustrated in Fig.~\ref{fig:algorithm_framework}, we maintain a lookup table $\mathcal{H}: \{0,1\}^{K\times K\times C} \to \mathbb{N}$ that maps each distinct user scheduling decision $\bf{B}$ to an index of a cached semantic model $\theta$. The caching scheme enables T-DRL to improve training efficiency by reducing the action space and eliminating redundant fine-tuning of the semantic model. At each learning round, given the current cache state, the DRL agent first updates the user scheduling $\bf{B}$ using PPO algorithm. Then, we query the lookup table to determine whether the corresponding semantic model needs to be updated. Based on the query result, two caching events may occur as follows:
\begin{itemize}
\item Cache Hit: If $\bf{B}$ is found in the cache, the associated pre-trained semantic model can be retrieved. Thus, the PPO agent only needs to optimize the RIS's phase shifts $\boldsymbol{\Phi}$ and compression ratios $\bm\beta$, conditioned on the fixed semantic model $\theta$ and user scheduling decision $\bf{B}$.
\item Cache Miss: If $\bf{B}$ is not present in the cache, the PPO agent first generates the remaining control variables $(\boldsymbol{\Phi}, \bm\beta)$. Then, we train the unified semantic model capable of handling a wide range of transmission patterns, and adapt it to the new scheduling decision $\bf{B}$ using Low-Rank Adaptation (LoRA). The resulting fine-tuned model $\theta$ and its corresponding user scheduling $\bf{B}$ are then added into the cache for future reuse.
\end{itemize}
% This pre-training strategy enhances the generalization ability of the semantic model while significantly reducing computational overhead. Specifically, during fine-tuning of the semantic model, the RIS reflection phase \({\bm \varphi}(t)\) is treated as a layer of a neural network and optimized using backpropagating. Since \({\bm \varphi}(t)\) is a high-dimensional variable, many configurations of reflection phases may correspond to low rewards, which can lead to inefficient exploration if trained directly with DRL. By optimizing \({\bm \varphi}(t)\) directly via the supervised signal, the resulting experience tuples, i.e., $<\bm{B}(t), {\bm \varphi}(t), \bm\beta(t)>$, can provide useful guidance for DRL, effectively initializing the distribution over the \({\bm \varphi}\) action space rather than learning from scratch. After fine-tuning, the DRL then optimizes the variable \(\bm\beta(t)\) based on \(\bm{B}(t)\) and \(\bm{\varphi}(t)\), as if we had truncated \(\bm{\varphi}(t)\) out of the DRL's decision. Then, the scheduling decision $\bm{B}(t)$ and its corresponding LoRA-adapted semantic model are stored together in a cache table $\mathcal{H}$ for future reuse.

\begin{figure}[t]
    \centering
    \includegraphics[width=0.45\textwidth]{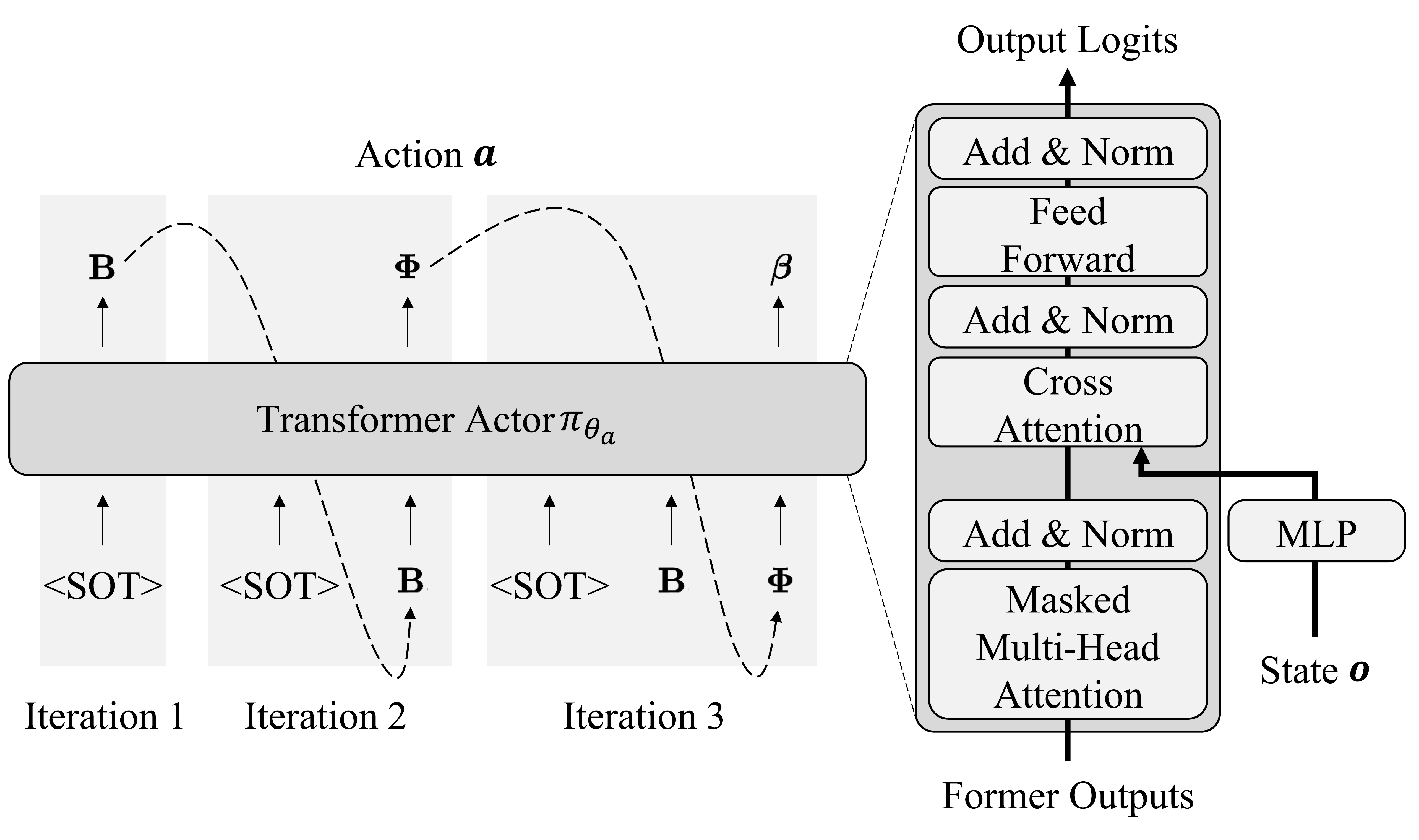}
    \caption{Transformer-based actor network for truncatable decision making.}
    \label{fig:variables embedding}
\end{figure}
\subsubsection{Transformer-based actor network}
Given the model caching strategy, the DRL agent no longer needs to optimize all control variables at every decision step. Instead, the variables with reduced action space are optimized depending on whether a cache hit or cache miss occurs. This dynamic action space requires the PPO actor network to produce action vectors of varying dimensionality.  However, conventional MLP-based actor networks are inherently limited to generating fixed-length action vectors, which makes them unsuitable for cache-aware decision-making. To address this limitation, we propose a transformer-based actor architecture that supports dynamic-length action generation. By exploiting the autoregressive property of the transformer, the actor is able to generate variable-length action outputs~\cite{Ding-2024IOTJ}.

As illustrated in Fig.~\ref{fig:variables embedding}, the action generation process encodes all control variables as binary vectors. We define a minimal binary vocabulary $\{0,1\}$ and encode each action as a sequence of tokens from this vocabulary. The current system state is first tokenized and embedded into a latent representation, which serves as the initial input to the actor network. The actor then generates the first action token autoregressively. This token is appended to the input sequence, and the extended sequence is fed back into the network to predict the next token.  By repeating this process, the actor sequentially produces an action sequence whose length can be flexibly controlled. Therefore, the action dimensionality can be dynamically adjusted according to different caching events.

The T-DRL framework is summarized in Algorithm~\ref{alg:DRL train1}. The PPO agent first determines the user scheduling decision $\bf{B}$, as shown in line~\ref{line:shedule}. The algorithm further decides whether the semantic model and the other control variables need to be fine-tuned, as described in lines~\ref{line:semantic training}-\ref{line:semantic training end}. Let $\mathcal{L}_{est}<\epsilon$ indicate that the semantic similarity estimator has converged, where $\epsilon$ is a predefined threshold. The semantic similarity is obtained either through actual semantic transmission or via the DNN-based estimator $\mathcal{R}_p$, depending on the estimator's convergence status as in lines~\ref{line:estimate}-\ref{line:estimate end}. The instantaneous  reward is computed according to the semantic similarity, and the corresponding transition tuples are stored in the replay buffer for PPO's training as in lines~\ref{line:ppo update}-\ref{line:ppo update end}. In lines~\ref{line: minibatch}-\ref{line: minibatch end}, the agent collects the interaction experiences through semantic transmissions, which are sampled for training the estimator $\mathcal{R}_p$.

{\iffalse\textcolor{blue}{To evaluate the computational complexity of the proposed T-DRL method, we first analyze the dimensionality of the action space.
In a conventional MLP-based actor network, the policy outputs a fixed-length continuous vector of dimension $D_{\text{conv}} = K^2 C + N + K$. The continuous outputs are subsequently quantized into discrete actions. Thus, the induced discrete action space has a cardinality of $2^{D_{\text{conv}}}$, which grows exponentially with the system dimensions. Thus, the overall computational complexity is evaluated as  $\mathcal{O}(K^2 C + N)$. }
In contrast, our method exploits the inherent structural constraints of the optimization variables to compress the action space significantly, akin to action branching architectures \cite{tavakoli2018action}. By encoding the scheduling matrix $\mathbf{B}$ via dual binary indicator vectors and employing logarithmic indexing for the RIS's phase shifts $\bm{\varphi}$—leveraging the discrete nature of practical RIS hardware \cite{Wu2020DiscreteIRS}—the required output dimension is reduced to $D_{\text{prop}} = 2KC + \lceil \log_2 N \rceil + 7K$. Consequently, the asymptotic complexity is improved to $\mathcal{O}(KC + \log N)$. This transformation yields an exponential reduction in the RIS-associated search space complexity and linearizes the complexity with respect to user density, thereby proving that the proposed auto-regressive actor network is computationally more efficient and tractable than traditional architectures.\fi}

\begin{algorithm}[t]

\caption{T-DRL for JSRE-based semantic transmissions}
\begin{algorithmic}[1]\label{alg:DRL train1}

\STATE Initialize the number of the user $K$, RIS's size $N$, the users' positions, and the DNN parameters.

\FOR{$v=1:V$}
\FOR{$t=1:T$}
\STATE PPO agent first adapts the user scheduling $\mathbf{B}$\label{line:shedule}
\IF{Cache Hit} \label{line:semantic training}
\STATE Retrieve the corresponding semantic model $\theta$
\ELSE
\STATE Fine-tune semantic model $\theta$ via back-propagation
\ENDIF
\STATE Update $\boldsymbol{\Phi}$ and $\bm\beta$ via PPO \label{line:semantic training end}
\IF{$\mathcal{L}_{est}>\epsilon$}\label{line:estimate}
\STATE Evaluate $\xi_{r,k,c}$ via real environment
\ELSE
\STATE Evaluate $\hat{\xi}_{r,k,c}$ via estimator $\mathcal{R}_p$
\ENDIF\label{line:estimate end}

\STATE Calculate instantaneous  reward $r$ \label{line:ppo update}
\STATE Record the transition to the next state
\STATE Store the transition tuple to replay buffer
\STATE Sample from replay buffer and update DRL networks
\ENDFOR \label{line:ppo update end}

\FOR{$e=1:E$}
\label{line: minibatch}
\STATE Sample mini-batches from replay buffer
\STATE Update estimator's parameters based on sampled data
\ENDFOR
\label{line: minibatch end}
\ENDFOR
\end{algorithmic}
\end{algorithm}
%\vspace{-0.1cm}

\section{Simulation Results \label{sec:results}}
%\vspace{-0.1cm}
In this section, we present numerical results to evaluate the performance of the JSRE scheme within RIS-assisted multi-user semantic communication systems.
%We set up a simulation environment with $K=5$ users and $T=5$ time slots. Under this severely time-slot-constrained condition, the superiority of multiple access techniques over TDMA is highlighted, and the spectral efficiency performance of various transmission schemes can be effectively compared.
The users are uniformly distributed within a circular area of radius $10$ meters, with the RIS positioned at the center to ensure coverage for all users. The Rician factor is set to $\kappa= 10$.  Other simulation parameters are detailed in Table~\ref{table:simulation_parameters}, similiar to those in~\cite{Yan2022SemRate4text}. The transmitted symbols are quantized using $16$-QAM.
The proposed JSRE framework is trained on a subset of ImageNet~\cite{Deng2009ImageNet} consisting of $310,000$ images, and the evaluation is performed on the Kodak24 dataset~\cite{kodak24}.

\begin{table}[h!]
\centering
\caption{Parameter settings in the simulation.}
\begin{tabular}{| m{4cm} | m{4cm} |}
\hline
\textbf{Parameters} & \textbf{Settings} \\
\hline
Number of users $K$ & $5$\\
Transmit power $p_d$ & $40$dBm \\
Background noise power $\sigma^2$ & $-174$dBm/Hz \\
Rician factor & 10\\
Pathloss model & $128.1+37.6\log(d)$ dB \\
Sub-band bandwidth $W_c$ & $20$MHz\\
\hline
\end{tabular}
\label{table:simulation_parameters}
\end{table}

We compare the JSRE scheme with four benchmark methods i.e., NOMASC~\cite{Li2023NOMA-Semantic}, DeepJSCC~\cite{Letafati2025JSCC4image}, DeepMA~\cite{Zhang2023DeepMA} and TDMA. Specifically, NOMASC employs a DenseNet-based semantic extraction, where the semantic encoding is used for data compression and simultaneous transmission is realized via NOMA. DeepJSCC proposes an encoding scheme consisting of convolutional modules. In this work, no channel information is incorporated into the encoding process. DeepMA integrates the multi-access method into the JSCC framework by training matched transmitters and receivers, where the training objective focuses on minimizing the semantic loss. The pairwise training in DeepMA inherently prevents unpaired encoders and decoders from correctly decoding, thereby enabling parallel transmission.
Finally, we use bit-based TDMA and semantic-based TDMA as the benchmark methods to highlight the necessity of the multiple access scheme. For bit-based TDMA, joint photographic experts group (JPEG) and low-density parity-check code (LDPC) are adopted as the source and channel coding methods, respectively. The LDPC code is configured with a block length of $n = 1296$ and a code rate of $R = 2/3$. As such, the number of bits per parity-check equation is $d_c = (1 - R)n = 432$ and the number of parity-check equations is $d_v = 144$.

\subsection{Illustrative Results for JSRE Transmission}

\begin{figure}
    \centering
    \subfigure[Coding efficiency comparison.]{
    \includegraphics[width=0.23\textwidth]{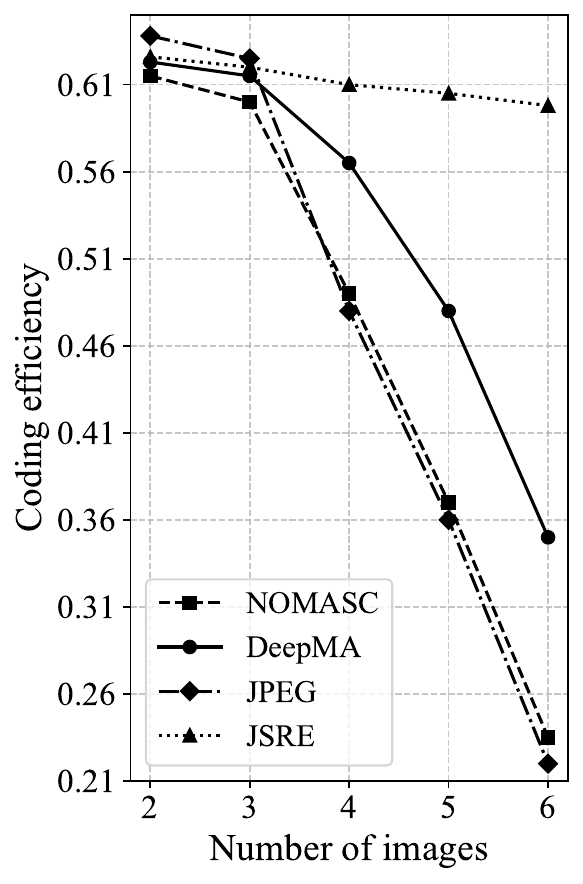}
    \label{fig:compression_efficiency}
    }
    \hspace{-0.6cm}
    \subfigure[SSIM comparison.]{
    \includegraphics[width=0.23\textwidth]{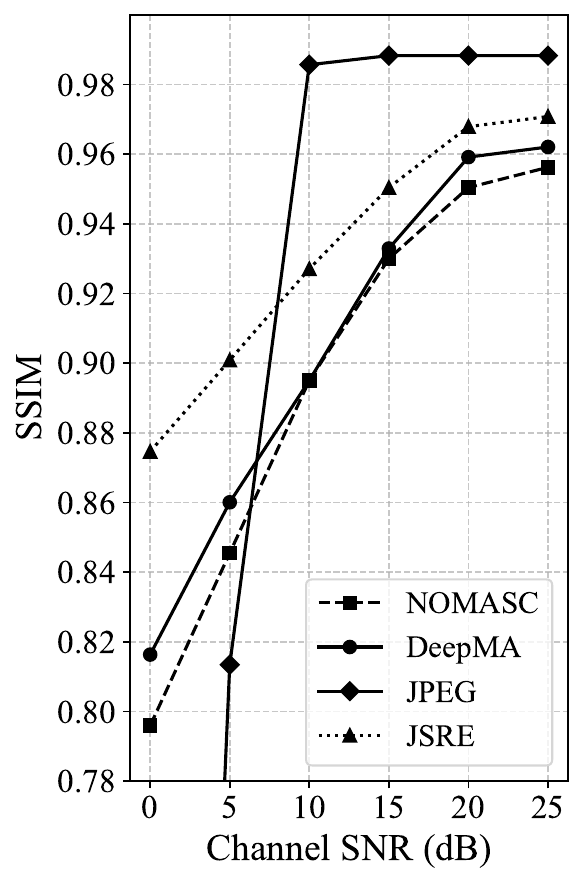}
    \label{fig:compression_SSIM}
    }
    \caption{The coding efficiency under various environmental settings.}
    % \vspace{-0.4cm}
\end{figure}

{
\iffalse
\begin{figure}
    \centering
    \includegraphics[width=0.45\textwidth]{figures/compression_efficiency.pdf}
    \vspace{-0.4cm}
    \caption{The coding efficiency performance with varying image numbers.}
    \label{fig:compression_efficiency}
\end{figure}

\begin{figure}
    \centering
    \includegraphics[width=0.45\textwidth]{figures/SNR_vs_PSNR_rebuild-from-csv_20250616_1642.pdf}
    \vspace{-0.4cm}
    \caption{The semantic recovery performance with varying channel SNR.}
    \label{fig:compression_SSIM}
\end{figure}
\fi}

\begin{figure}
    \centering
    \includegraphics[width=0.45\textwidth]{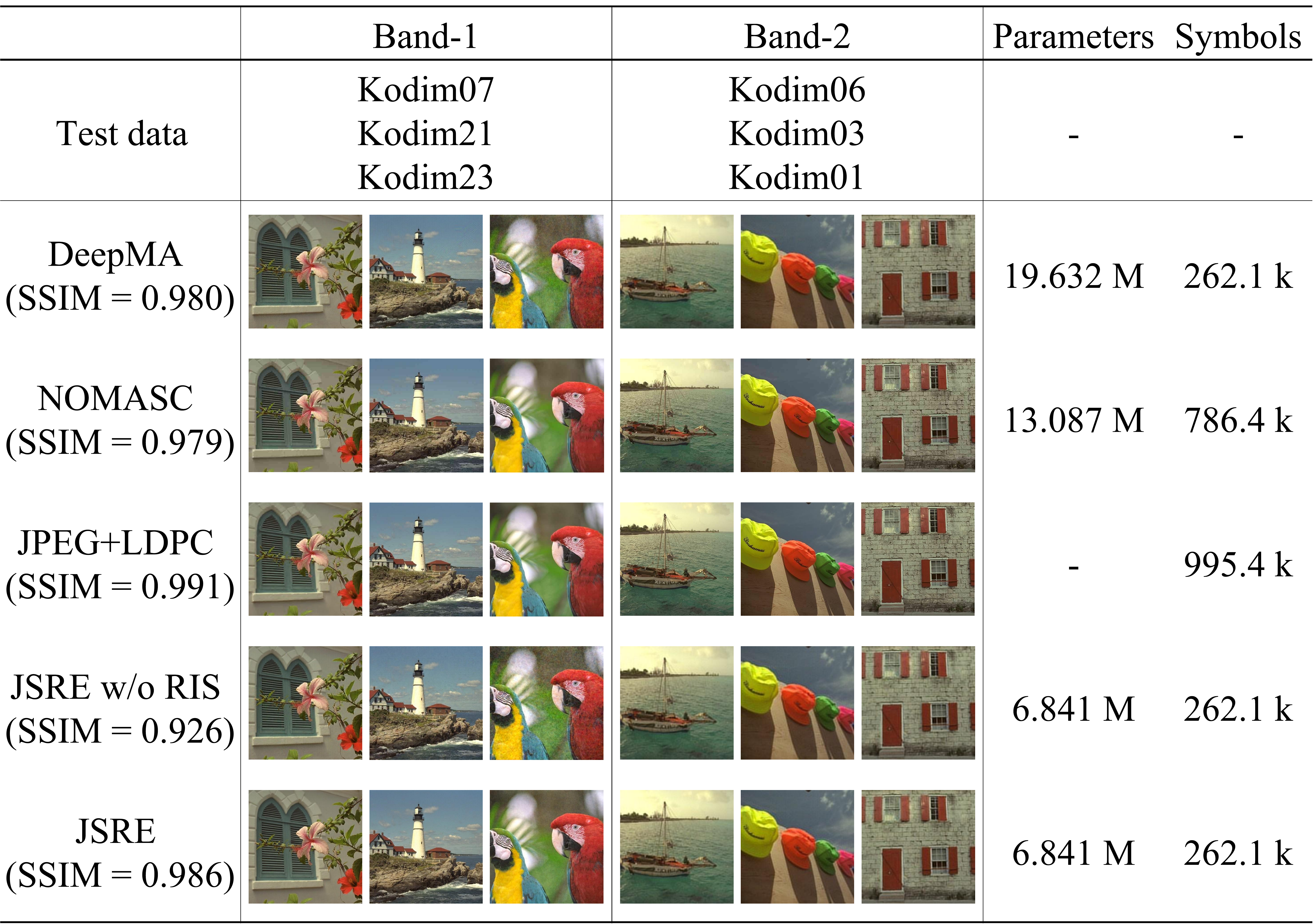}
    \caption{Visual reconstruction quality under the JSRE framework}% under slow fading channel. The proposed system transmits 3 pictures corresponding to 3 users simultaneously.}
    \label{fig:JSRE-Original_results}
\end{figure}
We first verify the effectiveness of the proposed JSRE framework. To provide a fair and intuitive comparison, a unified compression ratio is adopted for all benchmark algorithms.
As shown in Fig.~\ref{fig:compression_efficiency}, we define coding efficiency as the ratio of SSIM to traffic demand, and compare this metric across schemes as the number of users increases. It can be observed that JSRE and DeepMA achieve higher coding efficiency, and this advantage becomes more pronounced as the number of users grows. This is because bit-level schemes, such as JPEG and NOMASC, perform direct image compression with independent encoding and decoding for each image, which leads to higher transmission energy consumption when more users are involved.  These two methods do not employ encoding superposition. Consequently, their coding efficiency degrades linearly as the number of images increases.
In contrast, semantic-based schemes, i.e., JSRE and DeepMA, rely on semantic extraction to reduce the traffic demand, thereby substantially improving coding efficiency. Notably, by incorporating user-specific CSI into the decoding process, JSRE further enhances the orthogonality of multi-user semantic representations, thus achieving the highest coding efficiency among all schemes.

Figure~\ref{fig:compression_SSIM} illustrates the relationship between image reconstruction quality and the SNR. For the JSCC method, the algorithm adapts its encoding strategy to the instantaneous channel conditions, which leads to a smooth improvement in reconstruction quality as the SNR increases. In contrast, conventional separate source-channel coding is highly sensitive to channel errors. At low SNR, bit errors during transmission often cause decoding failures and severe degradation in reconstructed images. At high SNR levels, channel errors become negligible, and the reconstruction quality is determined mainly by the source coding scheme. In our experiments, JPEG2000 is used as the source codec with a fixed compression ratio. Although JPEG2000 supports both lossy and lossless modes, the configuration adopted here is lossy mode. Nevertheless, when the SNR is sufficiently high, for example, above $25$ dB, the performance of this traditional scheme approaches its practical limit because further increases in SNR no longer improve reconstruction quality. Thus, the performance of all methods saturates around $25$ dB, indicating that the system becomes limited by the source coding rather than the channel. Therefore, to ensure a fair and meaningful comparison under practical operating conditions, we fix the channel SNR at $25$ dB in all subsequent evaluations.

Figure~\ref{fig:JSRE-Original_results} illustrates the image reconstruction performance at the receiver under different methods. We consider an ideal scenario with relatively low background noise and light transmission load to explore the upper-bound performance of these methods. Specifically, two frequency bands are configured, and each band simultaneously transmits three RGB images with a resolution of $512\times512$. The channel SNR is set to $25$ dB, and transmission latency is neglected in this setting. We employ JPEG as a benchmark method for lossy image compression. Since JPEG operates independently of channel conditions, it can be regarded as providing an upper bound on reconstruction fidelity under ideal transmission scenarios without errors. In such cases, JPEG  theoretically outperforms semantic-based schemes in terms of reconstruction quality~\cite{Bourtsoulatze2019DeepJSCCImage}.
% Although JPEG adaptively adjusts its compression ratio according to image content, it still achieves a high compression efficiency on the Kodak24 dataset while maintaining near-lossless image reconstruction.}
% {\color{red} [Explain why JPEG is the best, why JSRE achieves performance closest to JPEG, and why the other benchmark methods is relatively poorly. The reasons are more important than results.]}
When energy efficiency and latency are not taken into account, JPEG achieves the best reconstruction performance among all considered methods, attaining an SSIM of $0.991$. The proposed JSRE scheme achieves an SSIM of $0.986$, which is comparable to that of JPEG. In contrast, removing the RIS component from JSRE leads to an SSIM degradation of $0.06$, clearly demonstrating the significant performance gain brought by the RIS-assisted channel reconfiguration.

\subsection{Convergence Performance of T-DRL Framework}
\begin{figure*}
    \subfigure[The training performance of the T-DRL framework.]{
        \includegraphics[width=0.33\textwidth]{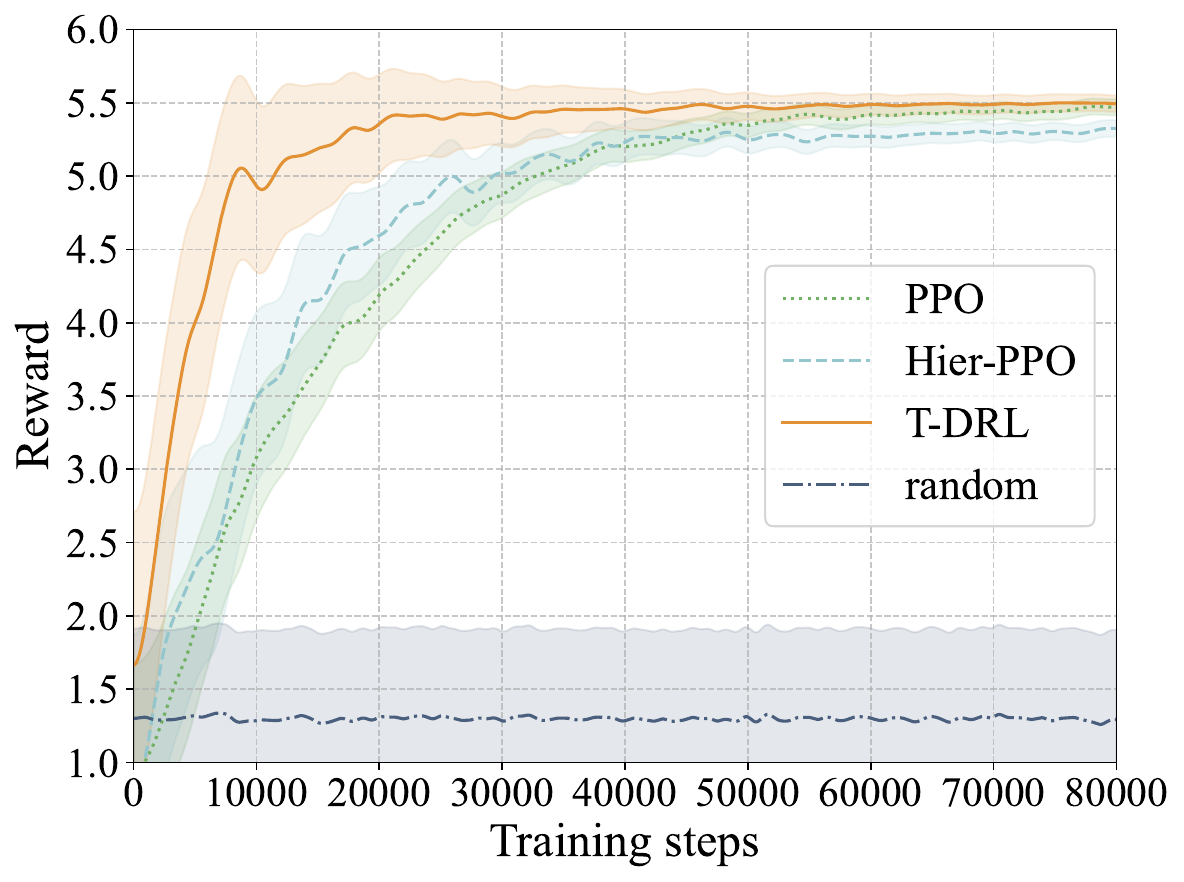}
        \label{fig:reward2}
    }
    \subfigure[Convergence under various RIS sizes.]{
    \includegraphics[width=0.33\textwidth]{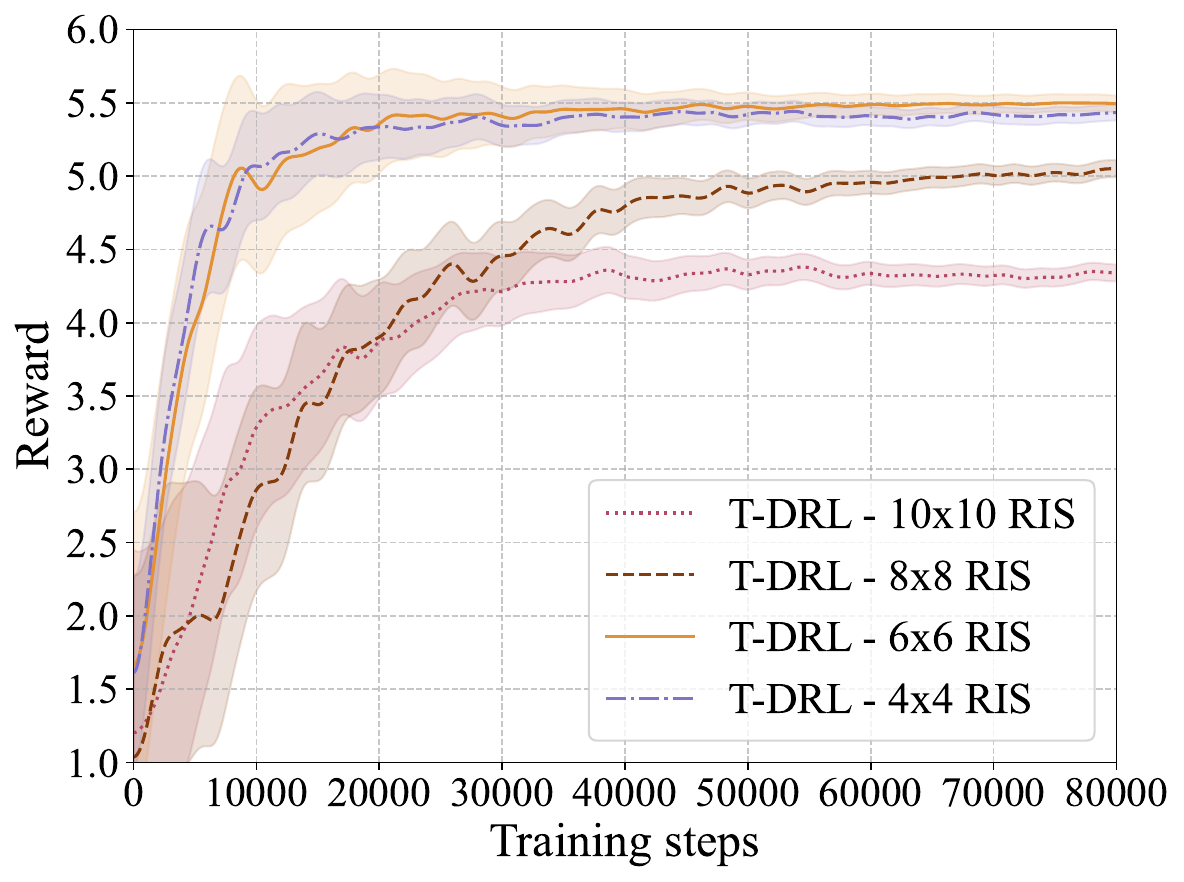}
    \label{fig:convergence_vs_RIS}
    }
    \subfigure[Convergence under different numbers of users.]{
    \includegraphics[width=0.33\textwidth]{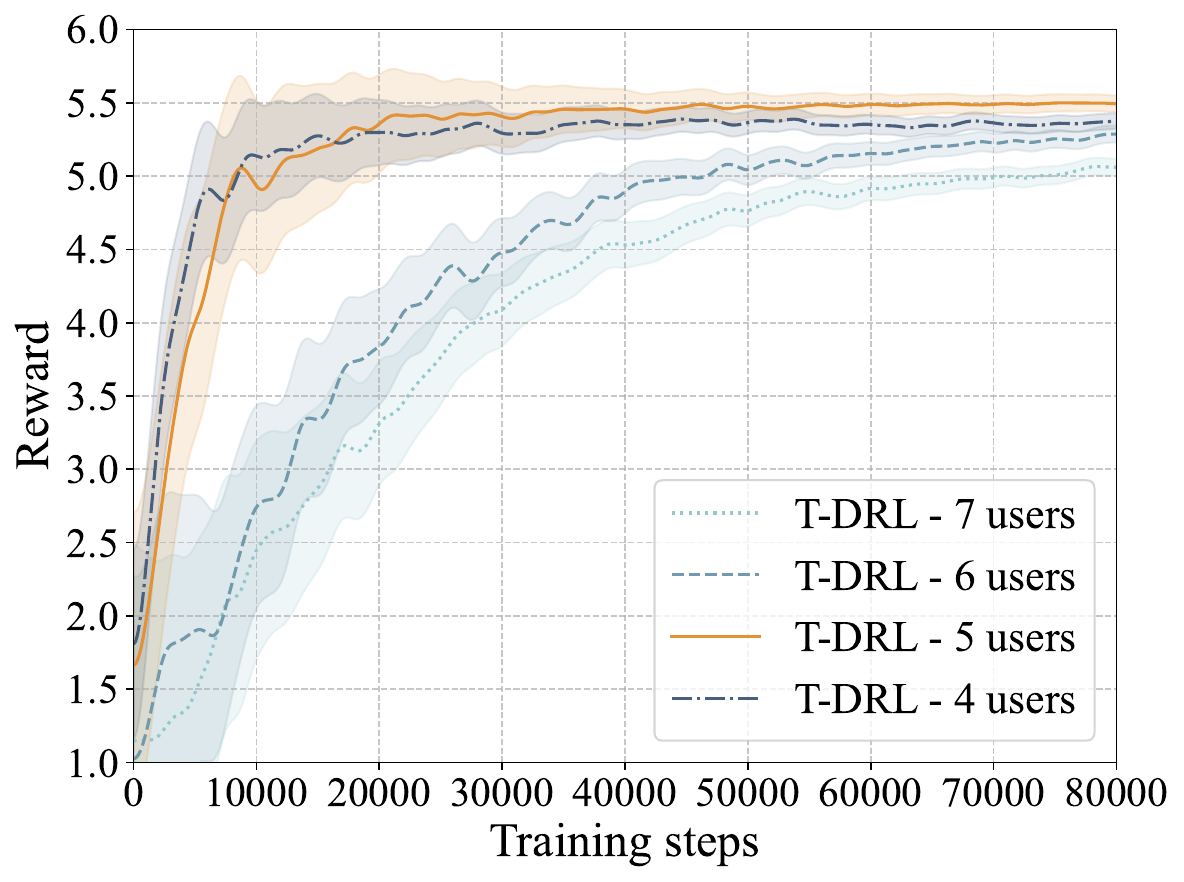}
    \label{fig:convergence_vs_user}
    }
    \caption{Convergence performance of different DRL and parameters}\label{fig:reward-all}
\end{figure*}
{\iffalse
\begin{figure}
    \centering
    \includegraphics[width=0.45\textwidth]{figures/weighted_IRSscale-6-copy__plot.pdf}
    \label{fig:reward2}
    \vspace{-0.4cm}
    \caption{The training performance of the T-DRL framework.}
    \label{fig:convergence}
\end{figure}
\fi}
In this part, we evaluate the learning performance of the T-DRL framework compared to benchmark learning algorithms: standard PPO and hierarchical PPO (denoted by Hier-PPO).
Figure~\ref{fig:reward-all}(a) shows the convergence behavior of these methods. The baseline PPO employs an end-to-end training paradigm, where the policy network directly maps the full system state to a complete action vector in a single forward pass, generating all optimization variables jointly. In contrast, Hier-PPO decomposes the decision-making process into multiple sub-policies organized across hierarchical levels. Each sub-policy is trained using reward signals propagated from higher-level modules, a design specifically introduced to mitigate the challenges posed by extremely large state and action spaces.
The convergence curves of T-DRL are generally faster than those of the baseline algorithms, albeit with slightly higher variance. This increased variance stems from the increased architectural complexity of the transformer-based actor network, which exacerbates optimization instability during the early stages of reinforcement learning exploration.

{\iffalse
\begin{table}
\centering
\caption{Comparison T-DRL variants.}
\label{tab:trunc-PPO_variants}
\begin{tabular}{lcccc}
\hline
Variants   & Duration & Converge Step($\times10^3$) & Energy Efficiency(suts/joule) & Variance\\
\hline
T-DRL & \textbf{6h 11m} & 21.19 & \textbf{5.52} & 0.07\\
T-DRL without action trunction  & 8h 33m & 36.14 & 5.27 & 0.07\\
T-DRL without estimator  & 9h 08m & \textbf{15.32} & 5.42 & 0.07\\
T-DRL without model cache  & 20h 18m & 15.54 & 5.50 & 0.07\\
\hline
\end{tabular}
\end{table}
\fi}
\begin{table}[htbp]
\centering
\caption{Comparison T-DRL variants.}
\label{tab:trunc-PPO_variants}
\newcolumntype{C}{>{\centering\arraybackslash}X}
\newcolumntype{L}{>{\raggedright\arraybackslash}X}

\begin{tabularx}{\columnwidth}{Lccc}
\hline
Variants & Duration & \begin{tabular}[c]{@{}c@{}}Converge \\step ($\times10^3$)\end{tabular} & \begin{tabular}[c]{@{}c@{}}Energy  \\efficiency (suts/J)\end{tabular}\\
\hline
T-DRL & \textbf{6h 11m} & 21.19 & \textbf{5.52} \\ \addlinespace[2pt]
T-DRL w/o  truncation & 8h 33m & 36.14 & 5.27  \\ \addlinespace[2pt]
T-DRL w/o estimator & 9h 08m & \textbf{15.32} & 5.42  \\ \addlinespace[2pt]
T-DRL w/o caching & 20h 18m & 15.54 & 5.50  \\
\hline
\end{tabularx}
\end{table}

{}

{Both the RIS configuration and the number of users directly affect the optimization complexity. In conventional DRL, this manifests as an increase in action dimensionality, which requires a wider output layer in the actor network. In contrast, T-DRL treats these variables as part of an input sequence. Thus, scaling the RIS size or user count merely extends the sequence length without fundamentally increasing the difficulty of the actor’s learning task. To validate this insight, Fig.~\ref{fig:convergence_vs_RIS} shows the convergence behavior of T-DRL under different RIS sizes. As the RIS size grows, T-DRL remains trainable and eventually converges, though with higher training variance and a slower convergence rate. The final performance also varies with RIS size: an excessively small RIS limits the system’s ability to shape the wireless environment, degrading image transmission quality and yielding lower rewards. Conversely, an overly large RIS incurs substantial energy overhead without delivering commensurate gains in semantic fidelity. A similar trend is observed when varying the number of users, as shown in Fig.~\ref{fig:reward-all}(c). Although T-DRL maintains convergence under increasing user numbers, its sequence-based architecture incurs higher training variance and slower convergence, reflecting a trade-off between scalability and learning efficiency.

Table~\ref{tab:trunc-PPO_variants} presents the ablation study on the T-DRL framework, quantifying the contribution of each component to training efficiency, convergence performance, and policy stability. All variants were trained for the same number of environment steps on an NVIDIA RTX 2080 Ti GPU, and the energy efficiency is reported in semantic units per joule (suts/J). The semantic unit was initially a concept defined in ~\cite{Yan2022SemRate4text} to distinguish it from the bit, as semantic communication focuses not only on the amount of data but also on the semantic similarity of the received data.  In this paper, we define the semantic unit as the reciprocal of the maximum modulation order achievable under an ideal channel.
The full T-DRL achieves a final energy efficiency of $5.52$ suts/J in $6$ hours and $11$ minutes of training. Removing the model cache drastically increases the training time to $20$ hours and $18$ minutes, with nearly identical performance. This confirms that the model cache is essential for alleviating the prohibitive computational overhead of re-evaluating the semantic transmission model at every training step. Moreover, action truncation also plays a critical role. When disabled, training takes longer and yields a lower energy efficiency of $5.27$ suts/J, along with increased policy variance.  This indicates that leveraging the cache to truncate the action space and guide the agent's exploration of the RIS's configuration based on cached optimal values is essential for faster convergence to a better final policy.
Finally, ablating the similarity estimator adds approximately $3$ hours to the training time, with only marginal changes in performance and variance. This validates the estimator’s effectiveness in accelerating environment interactions by replacing costly full-system simulations with fast, learned approximations.

\subsection{Energy Efficiency under Varying Network Scales}
In this section, we investigate the impact of key system configurations, i.e., the number of users and the size of the RIS, on the overall system performance. The number of users directly determines the transmission load, thereby influencing system energy efficiency, fairness among users, and computational complexity. Meanwhile, a larger RIS improves the quality of semantic encoding and decoding by enhancing channel conditions, but it also incurs higher hardware complexity and additional energy consumption. To understand these trade-offs, we analyze how variations in user number and RIS size affect the system performance in the following.
\begin{figure}
    \centering
    \subfigure[Energy efficiency.]{
    \includegraphics[width=0.23\textwidth]{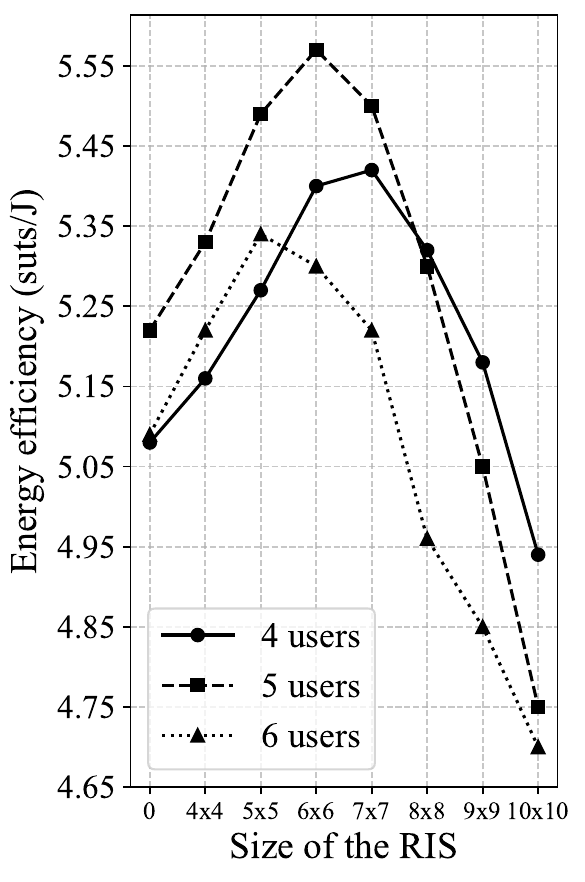}
    \label{fig:ris-vs-users}
    }
    \hspace{-0.65cm}
    \subfigure[SSIM performance.]{
    \includegraphics[width=0.23\textwidth]{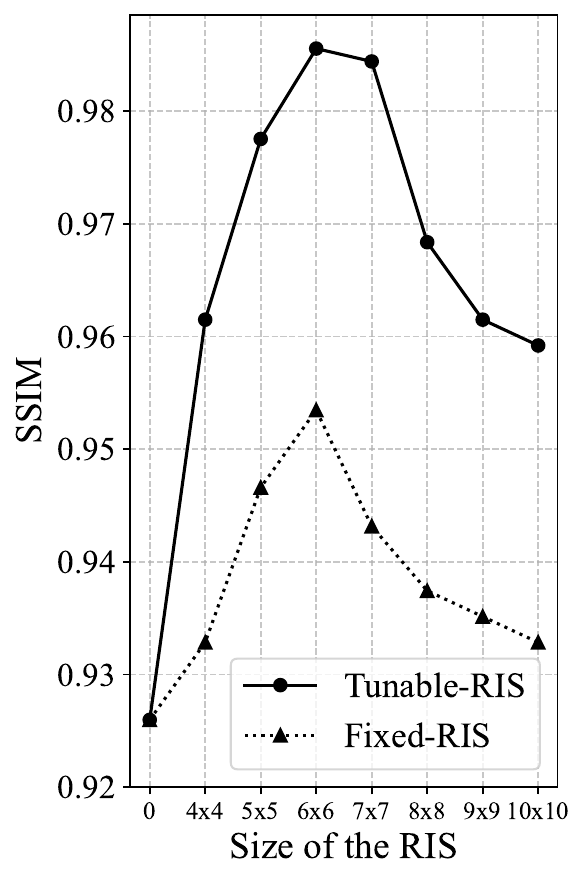}
    \label{fig:RIS-vs-PSNR}
    }
    \vspace{-0.2cm}
    \caption{System performance under different RIS sizes.}
\end{figure}

We first conduct a dedicated experiment to examine how RIS size influences system energy efficiency, as shown in Fig.~\ref{fig:ris-vs-users}. The system energy efficiency consistently follows a unimodal trend with respect to the RIS size. It first increases and then decreases as the RIS size increases. This behavior stems from the inherent characteristics of the discrete phase RIS. As the RIS size increases, the improvement in image transmission quality gradually saturates due to diminishing returns in channel conditioning. At the same time, the energy consumed by the RIS grows approximately quadratically with its size, primarily because of control circuitry and switching losses. As such, beyond a certain point, the marginal gain in transmission quality no longer compensates for the added energy cost, leading to a decline in overall energy efficiency. Moreover, the location of the peak energy efficiency and the corresponding optimal RIS size depend on the number of users. As the user number increases, the system faces higher transmission demand and greater multi-user concurrency, which intensifies inter-user interference. To cope with this, the policy tends to allocate fewer time slots per user or adopt higher compression ratios. Both strategies reduce semantic fidelity and degrade energy efficiency. Hence, the optimal RIS size shifts dynamically with the network load, highlighting the need for adaptive configuration in practical deployments.

{\iffalse
\begin{figure}
    \centering
    \includegraphics[width=0.45\textwidth]{figures/IRS_vs_PSNR_rebuild-from-csv_20250616_1641.pdf}
    \vspace{-0.4cm}
    \caption{The SSIM performance under varying RIS sizes.}
    \label{fig:RIS-vs-PSNR}
\end{figure}
\fi}

Figure~\ref{fig:RIS-vs-PSNR} examines the impact of RIS size on image transmission performance. The x axis represents the number of reflecting elements in the RIS, where a value of zero indicates the absence of an RIS. Two RIS configurations are evaluated: Tunable RIS, which uses optimized and trainable reflection coefficients, and Fixed RIS, which employs randomly initialized and non-adaptive coefficients.
The results show that for both configurations, SSIM improves with increasing RIS size compared to the baseline without an RIS. For Tunable RIS, performance increases sharply, reaching a maximum SSIM of approximately $0.987$ at a size of $6 \times 6$. Beyond this point, further enlargement of the RIS leads to a degradation in SSIM, suggesting that excessively large RIS arrays may be detrimental and that an optimal size exists for performance maximization. Similarly, Fixed RIS achieves its peak performance of about $0.954$ at a size of $ 6 \times 6$, followed by a decline at larger sizes. It is  also shown that although the peak image reconstruction quality is achieved at $6 \times 6$, the associated gain in SSIM does not compensate for the penalty incurred by the increased energy consumption of the larger RIS.

% \begin{figure}
%     \centering
%     \includegraphics[width=0.45\textwidth]{figures/barchart.pdf}
%     \vspace{-0.4cm}
%     \caption{The systems' FLOPs under different user numbers.}
%     \label{fig:users-vs-FLOPs}
% \end{figure}

% \begin{figure}
%     \centering
%     \includegraphics[width=0.45\textwidth]{figures/runtime.pdf}
%     \vspace{-0.4cm}
%     \caption{The user-wise energy efficiency under different user numbers.}
%     \label{fig:runtime}
% \end{figure}

\begin{figure}
    \centering
    \subfigure[The system's FLOPs.]{
    \includegraphics[width=0.23\textwidth]{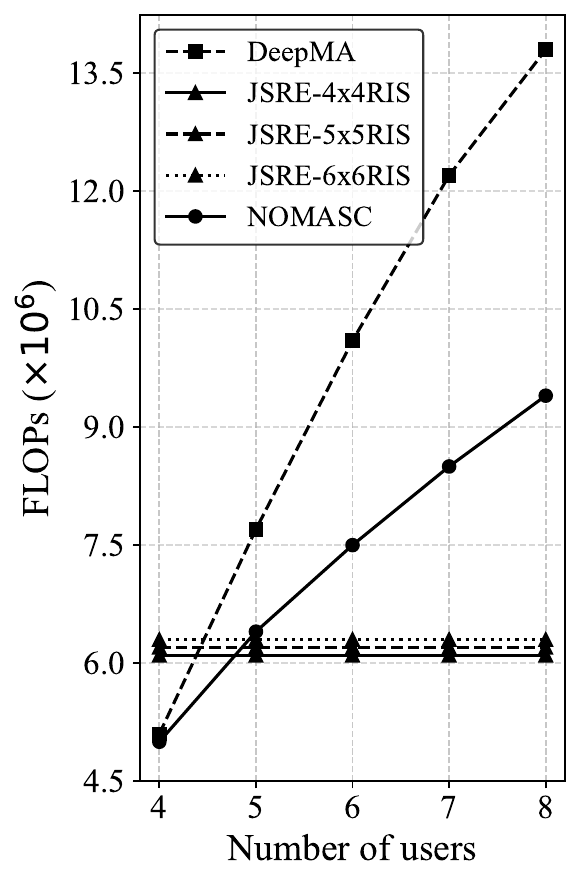}
    \label{fig:users-vs-FLOPs}
    }
    \hspace{-0.6cm}
    \subfigure[Average energy efficiency.]{
    \includegraphics[width=0.23\textwidth]{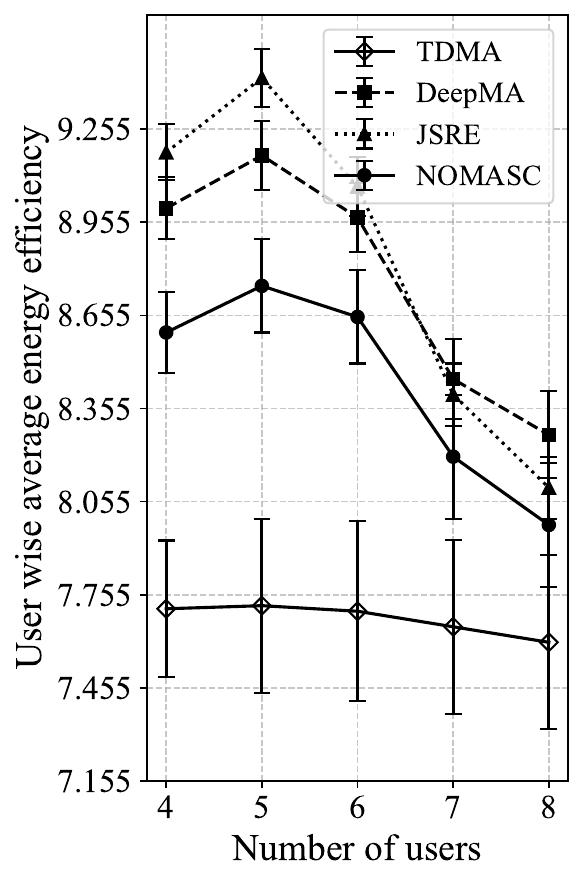}
    \label{fig:runtime}
    }
    \vspace{-0.2cm}
    \caption{Energy efficiency under different numbers of users.}
\end{figure}

Figure~\ref{fig:users-vs-FLOPs} compares the computational complexity of the learning based schemes in terms of FLOPs. The reported FLOPs reflect only the structural complexity of the models and are used to assess relative deployment costs. As such, FLOPs do not account for the actual computational load during inference.
It is seen that the computational cost of both NOMASC and DeepMA grows substantially with the number of users. DeepMA exhibits the steepest increase, with FLOPs rising from approximately $5.1$ million with four users to about $14$ million with eight users, making it the most computationally demanding scheme.  NOMASC also shows a significant rise, increasing from roughly $5.1$ million to $9.5$ million FLOPs over the same range. In contrast, the proposed JSRE method demonstrates superior computational efficiency and scalability, maintaining a nearly constant complexity of just above $6$ million FLOPs regardless of the user number.
In DeepMA, each user requires a dedicated semantic model. To support flexible scheduling among all users, every user must store not only its own model but also those of all other users. Consequently, the per-user model storage requirement scales linearly with the total number of users. NOMASC follows a similar principle, though it employs a lighter weight model, resulting in lower absolute overhead. JSRE unifies the semantic representation across all users and leverages channel state information to assist decoding, rather than relying on end-to-end user-specific training. Although its structural complexity is slightly higher than that of existing methods, it remains invariant as the number of users grows.  This  highlights the robust scalability of JSRE and underscores its advantage in multi-user transmission scenarios.

Figure~\ref{fig:runtime} presents a fairness performance analysis in terms of user-wise average energy efficiency, which is defined as the total system energy efficiency divided by the number of users. Overall, the curves for all schemes exhibit a unimodal trend: first increasing and then decreasing. This behavior arises because the system energy efficiency gradually saturates as the number of users grows, causing the per-user average to decline beyond a certain point. Among the evaluated methods, the proposed JSRE algorithm demonstrates superior fairness. It is observed that JSRE achieves significantly more consistent performance across users, particularly in large-scale scenarios. When the number of users is eight, the spread in user-wise average energy efficiency under JSRE is within $0.18$ suts/J. In comparison, DeepMA and NOMASC show larger disparities of approximately $0.3$ suts/J, while TDMA exhibits the poorest fairness, with a spread exceeding $0.6$ suts/J.

\section{Conclusion \label{sec:conclusion}}
In this paper, we have proposed a JSRE scheme for an RIS-assisted semantic communication system, enabling efficient simultaneous transmission among multiple users. To maximize semantic-aware system energy efficiency, we have introduced a T-DRL framework by jointly optimizing the user scheduling, RIS's phase shifts, and the semantic compression ratio.  The proposed JSRE algorithm unifies the semantic models across all users by leveraging user-specific CSI as a condition. This design exploits the distinct channel characteristics of each user to enhance the orthogonality of semantic representations, thereby reducing both model storage requirements and training overhead.  Furthermore, the T-DRL framework dynamically adjusts its action output dimension based on the current state of the model cache. It effectively lowers the dimensionality of the policy output, which  improves DRL's learning efficiency and accelerates convergence.  Numerical results demonstrated that the proposed JSRE scheme achieves significantly higher semantic-aware system energy efficiency compared to conventional benchmark schemes. These results also validated that incorporating user-specific CSI efficiently enhances the orthogonality of multi-user semantic signals, thus improving energy efficiency in concurrent semantic transmissions.

\ifCLASSOPTIONcaptionsoff
  \newpage
\fi
\vspace{-0.1cm}

\bibliographystyle{IEEEtran}
\bibliography{bibtex/bib/IEEEexample}

\end{document}